\documentclass[11pt, letterpaper]{article}
\usepackage{epsfig,amssymb,amsmath, multirow, url, tcolorbox,booktabs, enumitem}

\usepackage{hyperref}
\usepackage{graphicx}
\usepackage{enumitem}
\usepackage{natbib}
\usepackage[utf8]{inputenc}
\usepackage{authblk}
\usepackage{lscape}
\usepackage{adjustbox}
\newsavebox{\theorembox}
\newsavebox{\lemmabox}
\newsavebox{\claimbox}
\newsavebox{\factbox}
\newsavebox{\corollarybox}
\newsavebox{\examplebox}
\newsavebox{\remarkbox}
\newsavebox{\assbox}
\newsavebox{\propositionbox}
\newsavebox{\problembox}
\newsavebox{\defbox}

\savebox{\theorembox}{\noindent\bf Theorem}
\savebox{\lemmabox}{\noindent\bf Lemma}
\savebox{\factbox}{\noindent\bf Fact}
\savebox{\corollarybox}{\noindent\bf Corollary}
\savebox{\examplebox}{\noindent\bf Example}
\savebox{\assbox}{\noindent\bf Assumption}
\savebox{\propositionbox}{\noindent\bf Proposition}
\savebox{\problembox}{\noindent\bf Problem}
\savebox{\defbox}{\noindent\bf Definition}

\newtheorem{lem}{\usebox{\lemmabox}}

\newtheorem{@assumption}{\bf Assumption}[section]

 \newtheorem{@remark}{\bf Remark}[section]
 \newenvironment{remark}{\begin{@remark}\rm}{\end{@remark}}

\newcommand{\argmax}{\mathop{\rm argmax}}

\newcommand{\argmin}{\mathop{\rm argmin}}

\def\blot{\quad {$\vcenter{\vbox{\hrule height.4pt
             \hbox{\vrule width.4pt height.9ex \kern.9ex \vrule
width.4pt}
             \hrule height.4pt}}$}}

\newcommand{\doublebar}[1]{\bar{\bar{#1}}}
\newcommand{\doubletilde}[1]{\tilde{\tilde{#1}}}

\usepackage{multirow}
\providecommand{\keywords}[1]
{
{  \small	
  {\textit{Keywords---}} #1}
}
\usepackage{cleveref}
\crefname{equation}{equation}{equations}
\Crefname{equation}{Equation}{Equations}
\crefrangelabelformat{equation}{(#3#1#4--#5#2#6)}
\crefmultiformat{equation}{equations (#2#1#3}{, #2#1#3)}{#2#1#3}{#2#1#3}
\Crefmultiformat{equation}{Equations (#2#1#3}{, #2#1#3)}{#2#1#3}{#2#1#3}
\usepackage{caption}
\usepackage{subcaption}
\usepackage{graphicx}
\usepackage{rotating}
\usepackage{tikz}

\begin{document}
\title{Group integrative dynamic factor models \\ 
with application to multiple subject brain connectivity}

\author[1]{Younghoon Kim$^{*}$}
\author[2]{Zachary F. Fisher}
\author[3]{Vladas Pipiras} 

\affil[1]{Cornell University}
\affil[2]{The Pennsylvania State University}
\affil[3]{University of North Carolina at Chapel Hill}

\def\thefootnote{$*$}\footnotetext{Corresponding author. Email: yk748@cornell.edu}

\date{\today}
\maketitle

\begin{abstract}
    This work introduces a novel framework for dynamic factor model-based group-level analysis of multiple subjects time series data, called GRoup Integrative DYnamic factor (GRIDY) models. The framework identifies and characterizes inter-subject similarities and differences between two pre-determined groups by considering a combination of group spatial information and individual temporal dynamics. Furthermore, it enables the identification of intra-subject similarities and differences over time by employing different model configurations for each subject. Methodologically, the framework combines a novel principal angle-based rank selection algorithm and a non-iterative integrative analysis framework. Inspired by simultaneous component analysis, this approach also reconstructs identifiable latent factor series with flexible covariance structures. The performance of the GRIDY models is evaluated through simulations conducted under various scenarios. An application is also presented to compare resting-state functional MRI data collected from multiple subjects in autism spectrum disorder and control groups.
\end{abstract}



\section{Introduction}
\label{se:intro}

\subsection{Statistical analysis of brain connectivity}
\label{se:connectivity}

The analysis of brain connectivity typically focuses on examining the synchronized activity of spatially distinct brain regions. Modeling such interconnected regions includes structural connectivity (a network of anatomical or structural connections linking distinct units), and functional connectivity (capturing statistical dependence between distinct units). The units can be individual neurons, neuronal populations, and anatomical regions of interest (ROIs). In addition to the statistical dependence, one may be interested in causal interactions between these units (effective connectivity) \cite[see][for further explanation]{sporns:2007}.

To explore functional connectivity, one uses functional magnetic resonance imaging (fMRI) to capture blood-oxygen-level dependent (BOLD) signals observed over time. Each BOLD signal is mapped to a pre-specified ROI via a brain atlas. The strength of this signal is regarded as a proxy for brain activity since the blood oxygenation of the area of interest increases when the corresponding regions become active. Depending on the research goal, one may focus on establishing a statistical model to localize brain areas activated by the task, or building a prediction model about psychological or disease states. For these purposes, task-based fMRI (T-fMRI), a series of BOLD signals acquired while the subject performs a set of tasks, is often used. In making inferences about the structure of relationships among brain regions, across time points, and between subjects, resting-state fMRI (R-fMRI) is another option \cite[e.g.,][]{smith:2004}. Our motivating application will focus on R-fMRI.

Two types of questions can be addressed with R-fMRI for statistical brain network analysis. The first question concerns the construction of a network representation that reveals spatial and temporal dependences between ROIs. The most active research related to this question involves identifying commonality and uniqueness in connectivity patterns between (group of) subjects. The other question is about establishing a mathematical model that describes effective connectivity of the brain network  \cite[e.g.,][]{lindquist:2008}.

One promising statistical tool for addressing the first question is called data integration. In this context, data integration can be used to study similarities and differences in structures across subjects. Often, this heterogeneity is due to factors either directly observable (e.g., demographic, disease status, experiments) or hidden (e.g., causal mediation), which can be revealed through statistical modeling. The observations for such studies form multi-block (or multi-view) data, and these data structures naturally motivate the simultaneous exploration of the joint and (group) individual variations across and within data blocks. However, this approach often discounts temporal dependence and shows limitations in explaining directed connections. On the other hand, the second question can be dealt with using network modeling. For example, structural equation modeling (SEM) can be used to test causal relationships between ROIs. Alternatively, statistical dependence can be formulated through graphical models. However, identifying joint and (group) individual configurations in this context is less obvious, and the number of such studies is relatively low \cite[e.g.,][and references therein]{lee:2013}.


\subsection{Contributions}
\label{se:contribution}

We propose GRoup Integrative DYnamic factor models, which we abbreviate as GRIDY, and their estimation procedure for multivariate time series data (e.g., R-fMRI) collected on a set of common variables (e.g., ROIs) across subjects from several predetermined groups. There are several important points to note about GRIDY models. First, we can identify model configurations for spatial and temporal information that is shared or unique among subjects in the groups. Distinguishing model parameters contributes to explaining inter-subject differences in the measurement and dynamics of the time-dependent signals. Second, we allow the factor series to evolve over time. This is crucial because the model can present dynamic temporal dependencies, thus explaining intra-subject differences over time. Third, by allowing flexible covariance structures of factor series, the model can capture the heterogeneity of the subjects through different scales and combinations of factor series. Last, we illustrate the proposed model with an application to fMRI data from multiple subjects belonging to autism spectrum disorder and control groups. This work serves to suggest one possible choice of models to describe both inter-subject and intra-subject similarities and differences in dynamics of multiple subjects from several groups.


\subsection{Related approaches}
\label{se:related}

From a methodological standpoint, two developments are related to our proposed approach. One is integrative data analysis. It has been employed in various fields such as computational biology, chemometrics, and some areas of neuroscience. It originated from partial least squares (PLS) or canonical correlations analysis (CCA) to find the maximum covariance (correlation, respectively) between sample projections on the canonical loadings and the canonical scores from different subjects \cite[e.g.,][]{witten:2009,lofstedt:2013}. A variant of principal component analysis (PCA) for finding the direction presenting a maximal variation from the center has also been leveraged for multi-block data \cite[e.g.,][]{abdi:2013}.

The data integration approach has been developed to discover common and individual structures shared across different subjects. The recent development tends to include various block segmentations of multi-block data \cite[e.g.,][]{o:2019,park:2020}. Several data integration approaches have also been applied to neuroimaging studies. For example, \cite{yu:2017} analyzed CIFTI format imaging data (cortical surface plus subcortical gray matter) and a set of behavioral measurements from the Human Connectome Project (HCP) 500 subjects to understand the relationship between brain activity and human function. \cite{murden:2022} applied their model to R-fMRI and diffusion MRI (dMRI) to find similarities between functional connectivity and structural connectivity. The approaches above extract common structures from different datasets, with the heterogeneity arising from the source of data rather than factors uniquely embedded in subjects. These approaches have the limitation of disregarding temporal dependence.

Another related approach is Gaussian graphical models (GGMs). Several recent works on the joint estimation of these models incorporate temporal dependencies. For example, \cite{qiu:2016} used a kernel-based method to represent smooth changes in dependence over time. Similarly, \cite{qiao:2019} introduced Gaussian random functions to model temporal dependence, and \cite{zhu:2018} introduced matrix-valued variables to represent spatial and temporal dependence simultaneously. In contrast, \cite{fan:2018} added external covariates to loadings matrices in factor models to construct graphical models from the residuals. However, it is important to note that graphical models cannot be used to understand directional (causal) relationships between variables.

Our modeling approach is also closely related to various group-level analyses. In particular, a popular model in this framework is the group independent component analysis (GICA). GICA combines individual ICA models \cite[e.g.,][]{hyvarinen:2000,li:2007}, by using either a temporal or spatial concatenation of data across different subjects \cite[e.g.,][]{guo:2008}, or multi-way tensor models maximizing explained variations across temporal, spatial, and subject dimensions \cite[e.g.,][]{beckmann:2005}. A comprehensive survey regarding GICA models for group-level analysis can be found in \cite{calhoun:2009}. Other group-level analyses also exist \cite[e.g., population value decomposition in][]{crainiceanu:2011}. Closer to our proposed model, \cite{maneshi:2016} proposed a GICA model with shared and group specific components, called shared and specific independent component analysis (SSICA). However, the focus of these methods is not on factor dynamics over time, and very few approaches go beyond just identifying shared structures. In particular, the SSICA of \cite{maneshi:2016} involves several whitening/PCA-reduction steps at the group and aggregation levels on the temporal side whose effects are not easy to parse out.

Time series models offer an alternative as they easily incorporate directed and lagged relationships. Their usefulness in the analysis of fMRI has been demonstrated \cite[e.g.,][]{chen:2011,gates:2012}. Recently, \cite{skripnikov:2019}, \cite{manomaisaowapak:2022} employed combinations of different penalized estimations of vector autoregressive (VAR) models to represent subject heterogeneity by specifying both joint and individual structures. However, the joint structures in their works need not be identical, which has a limitation in explaining common dependences. \cite{fisher:2022,fisher:2023} addressed this issue by imposing additivity on the transition matrices, ensuring that the model produces identical estimates across subjects while allowing the rest of the components to be individually specified. However, VAR-based modeling requires a relatively large computational cost compared to other dimension reduction methods.

We shall employ dynamic factor models (DFMs) within the group-level analysis framework. DFMs provide low-dimensional dynamic representations that enhance explainability and typically involve low computational costs in their estimation procedure. Our study extends the application of DFMs to R-fMRI data from multiple subjects, enabling us to explore both inter- and intra-subject similarities and differences. Through the reconstruction of latent factor structures, we also investigate the heterogeneity among subjects. To the best of our knowledge, this work is the first model, even in terms of methodology, to consider DFMs in the context of multiple subjects and group-level analysis with shared and group specific structures.


\subsection{Outline of study}
\label{sse:outline}

The rest of the manuscript is organized as follows. In Section \ref{se:model}, we introduce our model consisting of joint (shared) and group individual structures. In Section \ref{se:method}, we outline the estimation procedure first, describe the segmentation of multi-block data and discuss the idenfiability of structures. Then, we introduce the estimation method for the loadings and factors in our model and the method to reconstruct the dynamics of factor series. In Section \ref{se:illustrative}, we provide simulation results under various settings that support the approach empirically. In Section \ref{se:data_app}, a joint analysis of R-fMRI data collected from multiple individuals is conducted and leads to the functional connectivity networks of brain regions. In Section \ref{se:discussion}, we summarize the results, discuss limitations, and suggest possible extensions.


\section{Proposed model}
\label{se:model}

Assume that we divide all individuals into $G$ non-overlapping and different groups and assume there are $K_1,K_2,\ldots,K_G$ number of subjects in each group. To simplify the notation and to be consistent with our application, we consider only two groups, $G=2$, and suppose that each group contains an equal number of subjects, with the first $K$ subjects in the first group and the remaining $K$ subjects in the second group. It is possible to extend the number of groups to be more than two and handle an unequal number of subjects in each group.

The observations from the $k$th subject are as follows. Define the $T_k\times d$ dimensional observation matrix $\mathbf{X}_{k}=(X_{i,t}^{(k)})_{i=1,\ldots,d,\ t=1,\ldots,T_k}$ to represent $k$th subject and $d$ variables with the observations collected over $T_k$ times. We assume that the temporal evolution of $k$th subject observations is explained by $r$ latent factor series, in the presence of the additive noise $\mathbf{E}_{k}\in\mathbb{R}^{T_k\times d}$. More specifically, the model is defined as the sum of the signals driven by $r$ latent factors and the additive noise, in a matrix form,
\begin{equation} \label{e:observation_k}
    \mathbf{X}_{k} = 
    \left\{\begin{array}{ll}
      \mathbf{F}_k\mathbf{B}_1'+\mathbf{E}_{k}, & \quad k=1,\ldots,K, \\
      \mathbf{F}_k\mathbf{B}_2'+\mathbf{E}_{k}, & \quad k=K+1,\ldots,2K,
    \end{array}\right.
\end{equation}
where $\mathbf{F}_k\in\mathbb{R}^{T_k \times r}$ are factor series of $k$th subject and the corresponding loadings matrices are denoted as $\mathbf{B}_1,\mathbf{B}_2\in\mathbb{R}^{d \times r}$ so that the components $\mathbf{F}_k\mathbf{B}_1'$ and $\mathbf{F}_k\mathbf{B}_2'$ construct the \textit{block signals} of $k$th subject in each group, respectively. 

We are interested in the separation of contributions for explaining the variability of the data jointly and group individually. Among the block signal of $k$th subject in $g$th group $\mathbf{F}_k\mathbf{B}_g'$, we assume that it can be decomposed into the joint components and group individual components, 
\begin{equation}\label{e:augmented_block1}
        \mathbf{F}_k\mathbf{B}_g' 
        = \bar{\mathbf{F}}_k \bar{\mathbf{B}}^{'} + \widetilde{\mathbf{F}}_k \widetilde{\mathbf{B}}_g^{'},
\end{equation}
where $\bar{\mathbf{B}}=(\bar{B}_{i,j})_{i=1,\ldots,d,\ j=1,\ldots,r_{J}}$ is the \textit{joint loadings matrix} identical to all $2K$ subjects, $\tilde{\mathbf{B}}_{g}=(\tilde{B}_{i,j}^{(g)})_{i=1,\ldots,d,\ j=1,\ldots,r_{G}}$, $g=1,2$, are \textit{group loadings matrices} only shared through the subjects within each group, $\bar{\mathbf{F}}_k \in \mathbb{R}^{T_k \times r_J}$ are \textit{joint factor series} and $\widetilde{\mathbf{F}}_k$ are \textit{group individual factor series}. We set $r=r_J+r_G$. Note that the ranks of the group individual factor series need not be identical; one can set the numbers of individual factors for groups 1 and 2 as $r_{G_1}\neq r_{G_2}$. However, to simplify the model description, we limit these quantities to be identical.

To construct DFMs, for each $k$, the $d-$dimensional vector of observation at time $t$, denoted as $X_{t}^{(k)}$, is explained by the joint component vector $\bar{\mathbf{B}} \bar{F}_{t}^{(k)}$ and group individual component vector $\tilde{\mathbf{B}}_{g}\tilde{F}_{t}^{(k)}$, $g=1,2$, where
\begin{eqnarray}
    X_{t}^{(k)} 
    &=& \mathbf{B}_{g}F_{t}^{(k)} + E_{t}^{(k)} \label{e:augmented_factor_1} \\
    &=& \begin{bmatrix} \bar{\mathbf{B}} & \tilde{\mathbf{B}}_{g} \end{bmatrix}
    \begin{bmatrix} \bar{F}_{t}^{(k)} \\ \tilde{F}_{t}^{(k)} \end{bmatrix} + E_{t}^{(k)} \\
    &=& \bar{\mathbf{B}} \bar{F}_{t}^{(k)} + \tilde{\mathbf{B}}_{g}\tilde{F}_{t}^{(k)} + E_{t}^{(k)} \label{e:augmented_factor_2}\\
    &=& \bar{X}_{t}^{(k)} + \tilde{X}_{t}^{(k)} + E_{t}^{(k)},\ g=1,2,\ k=1,\ldots,2K, \label{e:augmented_factor_3}
\end{eqnarray}
and $E_{t}^{(k)}=(E_{i,t}^{(k)})_{i=1,\ldots,d}$ is a noise vector with zero mean and variance $\boldsymbol{\Sigma}_{E,k}$. To connect \eqref{e:augmented_factor_1}--\eqref{e:augmented_factor_3} with \eqref{e:observation_k} and \eqref{e:augmented_block1}, we define $X_{t}^{(k)}$ as being explained by a linear combination of the $r_J-$dimensional joint factor $\bar{F}_{t}^{(k)}=(\bar{F}_{j,t}^{(k)})_{j=1,\ldots,r_{J}}$ and $r_G-$dimensional group individual factor $\tilde{F}_{t}^{(k)}=(\tilde{F}_{j,t}^{(k)})_{j=1,\ldots,r_{G}}$. In summary, the model is defined as
\begin{eqnarray} 
    \mathbf{X}_{k} &=& \begin{bmatrix}
                        X_1^{(k)'} \\
                        \vdots \\ 
                        X_{T_k}^{(k)'}
                        \end{bmatrix}  = \bar{\mathbf{X}}_{k} + \tilde{\mathbf{X}}_{k} + \mathbf{E}_{k}
                = \begin{bmatrix}
                        \bar{X}_1^{(k)'} \\
                        \vdots \\ 
                        \bar{X}_{T_k}^{(k)'}
                    \end{bmatrix} 
                + \begin{bmatrix}
                    \tilde{X}_1^{(k)'} \\
                     \vdots \\ 
                     \tilde{X}_{T_k}^{(k)'}
                \end{bmatrix}
               + \begin{bmatrix}
                    E_1^{(k)'} \\
                        \vdots \\ 
                    E_{T_k}^{(k)'}
                \end{bmatrix} \label{e:matrix_joint_individual1}  \\
                &=& \mathbf{F}_k\mathbf{B}_g'+\mathbf{E}_{k} 
            = \begin{bmatrix} 
                    F_1^{(k)'} \\
                    \vdots \\ 
                    F_{T_k}^{(k)'}
                \end{bmatrix}
                \begin{bmatrix}
                B_1^{(g)'} & \ldots &  B_{d}^{(g)'}
                \end{bmatrix}
                + \begin{bmatrix}
                E_1^{(k)'} \\
                \vdots \\ 
                E_{T_k}^{(k)'}
                \end{bmatrix}, \label{e:matrix_joint_individual2}
\end{eqnarray}
where the block signal for $k$th subject is 
\begin{equation}\label{e:augmented_block2}
        \mathbf{F}_k\mathbf{B}_g' 
        = \begin{bmatrix}
        \bar{F}_1^{(k)'} \\
        \vdots \\ 
        \bar{F}_{T}^{(k)'}
        \end{bmatrix}
        \begin{bmatrix}
        \bar{B}_1^{'} & \ldots &  \bar{B}_{d}^{'}
        \end{bmatrix} 
        + \begin{bmatrix}
        \tilde{F}_1^{(k)'} \\
        \vdots \\ 
        \tilde{F}_{T}^{(k)'}
        \end{bmatrix}
        \begin{bmatrix}
        \tilde{B}_1^{(g)'} & \ldots &  \widetilde{B}_{d}^{(g)'}
        \end{bmatrix}
        = \bar{\mathbf{F}}_k \bar{\mathbf{B}}^{'} + \widetilde{\mathbf{F}}_k \widetilde{\mathbf{B}}_g^{'}. 
\end{equation}

By stacking up all multi-block observations, we have a large matrix of blocks in \eqref{e:observation_k} as follows:
\begin{equation}\label{e:multi_block_AJIVE}
            \begin{bmatrix}
                    \mathbf{X}_1 \\
                    \vdots \\
                    \mathbf{X}_{2K} \\
                 \end{bmatrix} 
            = \begin{bmatrix}
                    \mathbf{Y}_1 \\
                    \vdots \\
                    \mathbf{Y}_{2K} \\
                 \end{bmatrix} 
+            \begin{bmatrix}
                    \mathbf{E}_1 \\
                    \vdots \\
                    \mathbf{E}_{2K} \\
                 \end{bmatrix}
              = \begin{bmatrix}
                    \bar{\mathbf{X}}_1 \\
                    \vdots \\
                    \bar{\mathbf{X}}_{2K} \\
                 \end{bmatrix} 
               + \begin{bmatrix}
                    \tilde{\mathbf{X}}_1 \\
                    \vdots \\
                    \tilde{\mathbf{X}}_{2K} \\
                 \end{bmatrix}
               + \begin{bmatrix}
                    \mathbf{E}_1 \\
                    \vdots \\
                    \mathbf{E}_{2K} \\
                 \end{bmatrix},
\end{equation}
where $\bar{\mathbf{X}}_k$ refers to the \textit{joint component} of $k$th subject and $\tilde{\mathbf{X}}_k$ refers to the \textit{group individual component} of $k$th subject, which form the joint structure and group individual structures, respectively. Each large matrix consists of $n=\sum_{k}T_k$ samples in time and $d$ variables, usually $n > d$. Moreover, since the model \eqref{e:multi_block_AJIVE} stacks up the observation blocks $\mathbf{X}_k\in\mathbb{R}^{T_k \times d}$ into a single matrix, unlike the tensor-based decomposition models where the sample lengths $T_k$ for all subjects should be equal, the sample length $T_k$ can be different across subjects. We also note that the way data blocks are stacked in \eqref{e:multi_block_AJIVE} is not the standard construction seen in the current literature on data integration. A typical stacking is to have subjects on the horizontal axis and variables across several datasets (blocks) on the vertical axis. The stacking used here arises because of the time dimension and is also tied to the model of interest. The number of blocks $2K$ in \eqref{e:multi_block_AJIVE} is also large, compared to what is typically considered in the literature on data integration.

To complete the construction of the latent factor series, assume that VAR($p$) model governs the joint factor series, that is,
\begin{equation}\label{e:filter_equation}
    \bar{\boldsymbol{\Psi}}_{k}(L)\bar{F}_{t}^{(k)} = \bar{\eta}_{t}^{(k)},\quad \left\{\bar{\eta}_{t}^{(k)}\right\} \sim \mathrm{WN}(0,\boldsymbol{\Sigma}_{\bar{\eta},k}),
\end{equation}
where $\bar{\boldsymbol{\Psi}}_{k}(L)=I - \bar{\boldsymbol{\Psi}}_{1,k}L - \ldots - \bar{\boldsymbol{\Psi}}_{p,k}L^p$ is a $r_J \times r_J$ operator of finite length $p$ with a lag operator $L$. The roots of $\textrm{det}(\bar{\boldsymbol{\Psi}}_{k}(z))$ lie outside the unit circle \cite[e.g. Chapter 2.1 in][]{lutkepohl2005new} so that the series becomes stable. Note that, unlike unspecified properties of the noise at the observation level, we quantify the noise for latent factor series to be white noise. The analogous VAR assumption is made for the group individual factor series.


\section{Methods}
\label{se:method}


\subsection{Overview of GRIDY procedure}
\label{sse:Procedure}

We summarize the estimation procedure of GRIDY. In the first step, by assuming that the block observations $\mathbf{X}_1,\ldots,\mathbf{X}_{2K}$ can be approximated by rank $r$ block signals as in \eqref{e:observation_k}, we obtain the joint components $\bar{\mathbf{X}}_1,\ldots,\bar{\mathbf{X}}_{2K}$ and two group individual components $\tilde{\mathbf{X}}_1,\ldots,\tilde{\mathbf{X}}_{K}$ and $\tilde{\mathbf{X}}_{K+1},\ldots,\tilde{\mathbf{X}}_{2K}$ as in \eqref{e:matrix_joint_individual1}--\eqref{e:matrix_joint_individual2} via a principal angle-based block segmentation algorithm. This step is explained in Section \ref{sse:segmentation}. Next, an algorithm that factorizes the obtained block structures into the loadings matrices $\bar{\mathbf{B}},\tilde{\mathbf{B}}_1,\tilde{\mathbf{B}}_2$ and the factor series $\bar{\mathbf{F}}_1,\ldots,\bar{\mathbf{F}}_{2K}$, $\tilde{\mathbf{F}}_1,\ldots,\tilde{\mathbf{F}}_{K}$, and $\tilde{\mathbf{F}}_{K+1},\ldots,\tilde{\mathbf{F}}_{2K}$ designed for multiple subject component models is applied separately to each block structure. Last, we reestimate the factor series from observation blocks by regression and complete the reconstruction of factor dynamics \eqref{e:filter_equation} by the Yule-Walker equations. The last two steps are explained in Section \ref{sse:reconstruction}. In addition, if $r$ is not explicitly given, we apply rotational bootstrap to estimate the unknown rank $r$ of the given block signals, which is described in Section \ref{ssse:initial_rank}. Note that each step of the proposed procedure can be replaced by some alternative method, some of which are considered in the simulation experiment in Section \ref{se:illustrative}.

\subsection{Segmentation of joint and group individual variation}
\label{sse:segmentation}

Joint Individual Variation Explained \cite[JIVE;][]{lock:2013} first used the idea of singular value decomposition (SVD) to segment concatenated block observations into joint and individual structures using low-rank approximations. In this approach, the desired structures are found by considering the sums of squared residuals to minimize the unexplained variability. However, a permutation-based rank selection for the joint structure is not guaranteed to achieve the true rank, in particular when there is some correlation across individual structures. Common orthogonal basis extraction \cite[COBE;][]{zhou:2015} can improve the basis extraction in JIVE in a systematic way by adding orthogonality constraints for joint and individual loadings. However, the algorithm is still iterative and the rank selection is not supported theoretically.

Recently, a non-iterative integrative data analysis framework, namely Angle-Based Joint Individual Variation Explained \cite[AJIVE;][]{feng:2018}, was developed. This framework decomposes multi-block data into the desired structures. Unlike previous works that determine the joint structure by enforcing orthogonality between the joint structure and the individual structure, AJIVE employs the concept of Principal Angle Analysis (PAA). Furthermore, the estimation procedure is non-iterative, which saves computational cost when the number of blocks is substantial. The details of the block segmentation algorithm and the identifiability condition of the produced block structures are presented in Sections \ref{ssse:estimation_structure} and \ref{ssse:identifiability}, respectively.

\subsubsection{Estimation of block structures}
\label{ssse:estimation_structure}

The AJIVE method applied to our problem works as follows. An initial rank estimate $\hat{r}_k$ of the signal block $\mathbf{Y}_k$ of $\mathbf{X}_k$ is needed for each $k$. Let
\begin{equation}\label{e:svd_subject}
    \hat{\mathbf{Y}}_k = \hat{\mathbf{U}}_k \hat{\mathbf{S}}_k \hat{\mathbf{V}}_k'
\end{equation}
be rank-$\hat{r}_k$ approximation of $\mathbf{Y}_k$ obtained from the SVD of $\mathbf{X}_k$, where $\hat{\mathbf{U}}_k$ and $\hat{\mathbf{V}}_k$ are $T_k \times \hat{r}_k$ and $d \times \hat{r}_k$ matrices of orthonormal columns and $\hat{\mathbf{S}}_k$ is a diagonal matrix with $\hat{r}_k$ singular values. Note that $\hat{\mathbf{V}}_k$ can be thought as an approximation of $\mathbf{B}_g$ under our model \eqref{e:matrix_joint_individual2}. Since we are looking for a joint structure among $2K$ subjects, it is natural to consider the following $( \sum \hat{r}_k ) \times d$ matrix and its SVD:
\begin{equation}\label{e:mat_J}
    \hat{J} 
    = \begin{bmatrix}
    \hat{\mathbf{V}}_1' \\ \vdots \\ \hat{\mathbf{V}}_{2K}' 
    \end{bmatrix} = \hat{\mathbf{U}}_J\hat{\mathbf{S}}_J\hat{\mathbf{V}}_J',
\end{equation}
where, in particular, the diagonal matrix $\hat{\mathbf{S}}_J$ consists of singular values $\hat{\sigma}_{J,i}$ in decreasing order. The idea then is to decide on the number $\hat{r}_J$ of singular values $\hat{\sigma}_{J,i}$ and take the corresponding columns of $\hat{\mathbf{V}}_J$ to capture a joint structure among $2K$ subjects.

Two approaches were suggested in \cite{feng:2018} for the number of singular values $\hat{\sigma}_{J,i}$ for a joint structure: the random direction bound approach and the Wedin bound approach. In the random direction approach, one replaces $\hat{\mathbf{V}}_{k}$ by randomly generated matrices with orthonormal columns and computes the largest singular value of the resulting matrix in \eqref{e:mat_J}. The process is repeated $M$ times and the 5th percentile of the distribution of such largest singular values is taken as a threshold for $\hat{\sigma}_{J,i}$. In the other approach, the Wedin bound (in the second inequality below concerning principal angles) is used to show that
\begin{equation}\label{e:singular_bound}
    \hat{\sigma}_{J,i}^2 \geq 2K - \sum_{k=1}^{2K} \sin^2(\theta_{k,r_k \wedge \hat{r}_k}) \geq 2K - \sum_{k=1}^{2K} \left\{ \frac{\max ( \|\mathbf{E}_k\hat{\mathbf{V}}_k\|_2, \|\mathbf{E}_k'\hat{\mathbf{U}}_k\|_2 )}{\sigma_{\min}(\hat{\mathbf{Y}}_k)} \wedge 1 \right \}^2,
\end{equation}
where $\sigma_{\min}(\hat{\mathbf{Y}}_k)$ denotes the smallest singular value of $\hat{\mathbf{Y}}_k$ in \eqref{e:svd_subject} and $\theta_{k,r_k \wedge \hat{r}_k}$ is the largest principal angle between subspaces for $\mathbf{Y}_k$ and $\hat{\mathbf{Y}}_k$. In practice, $\|\mathbf{E}_k\hat{\mathbf{V}}_k\|_2$ and $\|\mathbf{E}_k'\hat{\mathbf{U}}_k\|_2$ are approximated by $\|\mathbf{X}_k\tilde{\mathbf{V}}_k\|_2$ and $\|\mathbf{X}_k'\tilde{\mathbf{U}}_k\|_2$ where $\tilde{\mathbf{V}}_k$ and $\tilde{\mathbf{U}}_k$ have the same dimensions as and are orthogonal to $\hat{\mathbf{V}}_k$ and $\hat{\mathbf{U}}_k$, respectively. Repeating this process $L(=M)$ times, one obtains $L$ approximations of the lower bound in \eqref{e:singular_bound} and takes their 95 percentile as a threshold for the squared singular values $\hat{\sigma}_{J,i}^2$. The final determination of the joint rank $\hat{r}_{J}$ is achieved by counting the number of $\hat{\sigma}_{J,i}$ that exceed both thresholds.

Finally, define the matrix $\bar{\mathbf{V}}_J$ by retaining the $\hat{r}_J$ columns of $\hat{\mathbf{V}}_J$. Let $\bar{\mathbf{P}}_{J} = \bar{\mathbf{V}}_J\bar{\mathbf{V}}_J'$ be the projection matrix onto the estimated joint space. The estimators of $\bar{\mathbf{X}}_k$, the joint structure for each $k$, are obtained through $\mathbf{X}_k\bar{\mathbf{P}}_{J}$. Similarly, the estimation of the individual structure for each $k$, denoted as $\tilde{\mathbf{X}}_k$, is estimated by $\mathbf{X}_k(\hat{\mathbf{V}}_k\hat{\mathbf{V}}_k'-\bar{\mathbf{P}}_{J})$. We regard them as the estimators of group individual structures $\tilde{\mathbf{X}}_k$ in GRIDY model, but taking into account the following remark.

\begin{remark}\label{rem:rank}
We note that the joint structures $\mathbf{X}_k\bar{\mathbf{P}}_{J} = (\mathbf{X}_k\bar{\mathbf{V}}_J)\bar{\mathbf{V}}_J'$ are of the factor model form $\bar{\mathbf{F}}_k\bar{\mathbf{B}}'$ as in \eqref{e:augmented_block2}. This is not the case for the group individual structures $\mathbf{X}_k(\hat{\mathbf{V}}_k\hat{\mathbf{V}}_k' - \bar{\mathbf{P}}_J)$ where $k=1,\ldots,K$ for group 1 and $k=K+1,\ldots,2K$ for group 2. We nevertheless refer to the latter as group individual structures, and under our model, fit factor models of the form $\tilde{\mathbf{F}}_k\tilde{\mathbf{B}}_g'$ as in  \eqref{e:augmented_block2} through simultaneous component analysis (SCA) in Section \ref{ssse:estimation_factor} below. The factor model is also refitted through SCA for the joint structures. We also note that in our simulation study, we consider the procedure where the joint and group individual structures are identified through double application of SCA, but this approach performs considerably worse than the suggested method based on AJIVE. It should also be stressed that only AJIVE can currently produce reasonable estimates of the various ranks needed for our model. For further discussion, see Section \ref{se:illustrative}.
\end{remark}

\subsubsection{Identifiability of the block structure}
\label{ssse:identifiability}

Note that the model \eqref{e:augmented_factor_1}--\eqref{e:augmented_factor_3} should ideally satisfy two kinds of identifiability conditions. The first condition is to ensure that the factor model in \eqref{e:augmented_factor_1} can be uniquely decomposed into the joint components $\bar{\mathbf{F}}_{k}\bar{\mathbf{B}}'$ and the group individual components $\tilde{\mathbf{F}}_{k}\tilde{\mathbf{B}}_g^{'}$ as in \eqref{e:augmented_factor_2}. To distinguish this condition from the next one, we call it an identifiability condition. Another important condition is to ensure that the loadings matrices $\bar{\mathbf{B}},\tilde{\mathbf{B}}_1,\tilde{\mathbf{B}}_{2}$ and the factor series $\bar{F}_{t}^{(k)},\tilde{F}_{t}^{(k)}$ can also be identified. We will see that unlike the first condition, these loadings matrices and factor scores are not identifiable until additional assumptions are imposed. We shall refer to determinacy as the condition for identifying the loadings and the factor series in the components of our decomposition. This condition will be discussed in Section \ref{ssse:determinacy}.

The identification of the first two block structures in \eqref{e:multi_block_AJIVE} can be obtained by checking suitable identifiability conditions of AJIVE. The following lemma adapted from \cite{feng:2018} to our setting shows that under certain conditions, the uniqueness of the decomposed blocks can be achieved. To discuss the condition below, we define a subspace spanned by joint loadings matrix as 
\begin{equation*}
    \textrm{row}(\bar{\mathbf{X}}_{1}) = \ldots = \textrm{row}(\bar{\mathbf{X}}_{2K}) =: \textrm{row}(\bar{\mathbf{X}}).
\end{equation*}

\begin{lem}[\cite{feng:2018}]\label{lem:feng1}
Given a set $\{\mathbf{Y}_1,\ldots,\mathbf{Y}_{2K}\}$ of matrices, there are unique sets $\{\bar{\mathbf{X}}_1,\ldots,\bar{\mathbf{X}}_{2K}\}$ and $\{\tilde{\mathbf{X}}_1,\ldots,\tilde{\mathbf{X}}_{2K}\}$ of matrices so that
\begin{enumerate}
    \item $\mathbf{Y}_k = \bar{\mathbf{X}}_k + \tilde{\mathbf{X}}_k$ for all $k=1,\ldots,2K$.
    
    \item $\mathrm{row}(\bar{\mathbf{X}}_k)=\mathrm{row}(\bar{\mathbf{X}}) \subset \mathrm{row}(\mathbf{Y}_k)$ for all $k=1,\ldots,2K$.
    
    \item $\mathrm{row}(\bar{\mathbf{X}}) \perp \mathrm{row}(\tilde{\mathbf{X}}_k)$ for all $k=1,\ldots,2K$.
    
    \item $\bigcap_{k=1}^{2K} \mathrm{row}(\tilde{\mathbf{X}}_k)=\{\mathbf{0}\}$.
\end{enumerate}
\end{lem}

Condition 1 corresponds to our setting of interest. Note that each row space $\mathrm{row}(\bar{\mathbf{X}}_k)$ is defined by the loadings, specifically by the linear combinations of entries of loadings matrices multiplied by the factor scores. Also, note that the row subspaces of the joint components are spanned by the same loadings matrix, that is, $\textrm{row}(\bar{\mathbf{X}}_{k}) = \textrm{row}(\bar{\mathbf{F}}_{k}\bar{\mathbf{B}}') = \textrm{row}(\bar{\mathbf{B}}') = \textrm{row}(\bar{\mathbf{X}})$ since the subspace is represented by a linear combination of basis columns in the loadings matrix, where the coefficients are the factor scores, which rarely have the same values across different rows. Hence, condition 2 is satisfied assuming $\bar{\mathbf{X}}_k = \bar{\mathbf{F}}_k\bar{\mathbf{B}}'$ in our setting. Consider next condition 3. It assumes that for fixed $k$, $\bar{\mathbf{X}}_k \tilde{\mathbf{X}}_k^{'} = \bar{\mathbf{F}}_k\bar{\mathbf{B}}'\tilde{\mathbf{B}}_g\tilde{\mathbf{F}}_k' = \mathbf{0}_{T_k \times T_k}$, $g=1,2$, and is implied by $\bar{\mathbf{B}}'\tilde{\mathbf{B}}_1 = \bar{\mathbf{B}}'\tilde{\mathbf{B}}_2 = \mathbf{0}_{r_J \times r_G}$. This means that all columns in both of group loadings matrices are perpendicular to the corresponding columns of joint loadings matrix. The easiest way to achieve this condition is when the non-zero row entries of joint loadings matrix $\bar{\mathbf{B}}$ do not overlap with any other non-zero row entries of both group loadings matrices $\tilde{\mathbf{B}}_g$, $g=1,2$. Interestingly, this condition implies that the rows can be shared between group loadings matrices. Lastly, condition 4 states that there are no rows shared by all group loadings matrices. This is implied by our model where 
$\mathrm{row}(\tilde{\mathbf{B}}_1')\cap\mathrm{row}(\tilde{\mathbf{B}}_2')=\{0\}$. That is, the blocks span different subspaces depending on groups. We thus identified natural condition on the elements of our model $(\bar{\mathbf{B}},\tilde{\mathbf{B}}_1,\tilde{\mathbf{B}}_{2})$ so that its components are the unique components of the decomposition given in Lemma \ref{lem:feng1}.

We also note that the identifiability condition can be summarized as 
\begin{equation*}
    \mathbf{B}_g'\mathbf{B}_g 
    = \begin{pmatrix}
    \bar{\mathbf{B}}'\bar{\mathbf{B}} & \mathbf{0}_{r_J \times r_G } \\
    \mathbf{0}_{r_G \times r_J } & \tilde{\mathbf{B}}_g'\tilde{\mathbf{B}}_g \\
    \end{pmatrix}, \quad g=1,2.
\end{equation*}
As we will discuss in Section \ref{sse:reconstruction}, consideration of conditions on loadings matrices is not important to obtain the determinacy of factor series.



\subsection{Reconstruction of loadings matrices and factor series}
\label{sse:reconstruction}

After the segmentation using AJIVE, the remaining task is to reconstruct the factor structure in \eqref{e:augmented_factor_2}. A typical way to ensure the determinacy of the loadings matrices and factor series up to the transformation with the sign changes of the columns is to assume that the factor covariance is an identity matrix \cite[e.g.,][]{lam:2011,bai:2013}. For fixed $k$, by assuming we have only $k$th subject, the condition implies that the variances within the joint factor series and the group individual factor series should be identity matrices, 
\begin{equation}\label{e:orthogonality_factor} 
    \mathbb{E}\bar{F}_{t}^{(k)}\bar{F}_{t}^{(k)'} 
    = \mathbf{I}_{r_{J}}, \quad 
    \mathbb{E}\tilde{F}_{t}^{(k)}\tilde{F}_{t}^{(k)'} 
    = \mathbf{I}_{r_{G}}.
\end{equation}
The same determinacy condition has been applied for multiple subjects \cite[e.g.,][]{fan:2018} as well. This condition, however, is quite stringent; it does not allow for the joint or group individual factor series to be correlated and forces the covariance of factors to be identical across subjects, which limits the degree of heterogeneity.

To relax the stringent restrictions \eqref{e:orthogonality_factor}, we adopt the SCA introduced by \cite{timmerman:2003}. SCA was designed for multivariate time series from multiple subjects to study both intra- and inter-individual differences. We will demonstrate that the factor dynamics can also be reconstructed using this determinacy condition.

The idea of SCA is not different from PCA but various restrictions on the component factor scores may be imposed. There are four types of covariance structures suggested in \cite{timmerman:2003}. However, we focus on  SCA with PARAFAC2 constraints (SCA-PF2), inspired by tensor decomposition PARAFAC \citep[e.g.,][]{kiers:1999}. Additionally, we consider a more restrictive but useful covariance structure, namely SCA with INDSCAL constraints (SCA-IND). To compare the various covariance structures of the factors, consider the joint factor series. Then, the two covariance structures mentioned above are
\begin{eqnarray}
    \mathbb{E}\bar{F}_{t}^{(k)}\bar{F}_{t}^{(k)'} 
    &=:& \mathbb{E}\left(\bar{\mathbf{C}}_k\bar{A}_{t}^{(k)}\right)\left(\bar{\mathbf{C}}_k\bar{A}_{t}^{(k)}\right)^{'}  
    = \bar{\mathbf{C}}_k \bar{\boldsymbol{\Phi}} \bar{\mathbf{C}}_k, \quad \textrm{(SCA-PF2)} \label{e:SCA-PF2} \\
    \mathbb{E}\bar{F}_{t}^{(k)}\bar{F}_{t}^{(k)'} 
    &=:& \mathbb{E}\left(\bar{\mathbf{C}}_k\bar{A}_{t}^{(k)}\right)\left(\bar{\mathbf{C}}_k\bar{A}_{t}^{(k)}\right)^{'}  
    = \bar{\mathbf{C}}_k^2, \quad \textrm{(SCA-IND)} \label{e:SCA-IND}
\end{eqnarray}
where $\bar{\boldsymbol{\Phi}}=(\bar{\phi}_{i,j})$ is a $r_J \times r_J$ positive definite matrix and $\bar{\mathbf{C}}_k=(\bar{C}_{i,j}^{(k)})$ are diagonal matrices. Note that to identify matrices within their multiplications, we let the diagonal entries of $\bar{\boldsymbol{\Phi}}$ be unit-scaled so that they become correlation matrices of the scaled joint factors $\bar{A}_{t}^{(k)}$ and the absolute values of $\bar{\mathbf{C}}_k$ represent the standard deviations of the joint factors $\bar{F}_{t}^{(k)}$. Note also that \eqref{e:SCA-PF2} allows for correlations across joint factors while \eqref{e:SCA-IND} does not. Lastly, one can think of \eqref{e:SCA-IND} as a special case of \eqref{e:SCA-PF2} by letting $\bar{\boldsymbol{\Phi}}=\mathbf{I}_{r_{J}}$. Obviously, these conditions do not require any restriction on loadings matrices. The same assumption on the covariance of group individual factors can be applied, which in turn produces 
\begin{eqnarray*}
    \mathbb{E}\tilde{F}_{t}^{(k)}\tilde{F}_{t}^{(k)'} 
    &=:& 
    \tilde{\mathbf{C}}_k \tilde{\boldsymbol{\Phi}} \tilde{\mathbf{C}}_k, \\
    \mathbb{E}\tilde{F}_{t}^{(k)}\tilde{F}_{t}^{(k)'} 
    &=:&  
    \tilde{\mathbf{C}}_k^2,
\end{eqnarray*}
where $\tilde{\boldsymbol{\Phi}}$ is $r_G \times r_G$ positive definite and $\tilde{\mathbf{C}}_k=(\tilde{C}_{i,j}^{(k)})$ are diagonal as similarly defined as above. The details of the fitting algorithm and the determinacy of the factor series are in Sections \ref{ssse:estimation_factor} and \ref{ssse:determinacy}, respectively.

\begin{remark}
While the conditions as in \eqref{e:SCA-PF2}--\eqref{e:SCA-IND} enable the model to have more complex covariance structures, one can think that with more restrictions, as \eqref{e:orthogonality_factor} in an extreme case, the model becomes more parsimonious. In the other direction, the most flexible covariance structure is known as SCA with Invariant Pattern (SCA-P) \cite[see][]{timmerman:2003}. To our best knowledge, there are no tests to decide on particular structures. As long as we deal with the latent factor series and their characteristics are not known in advance, it is beneficial to allow the model to be as flexible as possible. Therefore, we do not prefer the most restrictive condition \eqref{e:orthogonality_factor}. At the same time, our hope is to incorporate the heterogeneity of subjects without losing identifiability. However, the SCA-P condition is known not to guarantee such uniqueness. Therefore, we prefer the SCA-PF2 condition and as a special case, the SCA-IND condition.
\end{remark}

\begin{remark}
Among studies related to SCA, using a rotational matrix to make targeted factor scores \cite[e.g.,][]{schouteden:2013,schouteden:2014} or particular structural loadings matrices \cite[e.g.,][]{helwig:2017} has been tried. However, the former requires a known target structure of factor scores in advance or requires additional steps to determine the target. It also carries relatively large computational costs since estimation of factor scores and loadings are involved in finding rotational transformation. Furthermore, the adopted determinacy condition in the approach is SCA-P, which does not guarantee the identification of factor scores. The latter also demands a pre-specified structure of loadings matrices in advance. When we do not know the relationship between factors themselves or desired loadings matrices, this blocking approach has limitations. 
\end{remark}

We estimate loadings matrices $\bar{\mathbf{B}},\tilde{\mathbf{B}}_1,\tilde{\mathbf{B}}_2$ and factor series $\{\bar{\mathbf{F}}_k,\tilde{\mathbf{F}}_k\}_{k=1,\ldots,2K}$ through a direct fitting algorithm for SCA-PF2 model, which will be described separately in Section \ref{ssse:estimation_factor}. As the last step of the estimation procedure, we refit the latent factors through subject-wise linear regressions:
\begin{equation}\label{e:refitting}
    \begin{pmatrix} \hat{\bar{\mathbf{F}}}_k & \hat{\tilde{\mathbf{F}}}_k \end{pmatrix} 
    = \argmin_{\mathbf{F}_k=(\bar{\mathbf{F}}_k\ \tilde{\mathbf{F}}_k)} 
    \| \mathbf{X}_k 
    - \bar{\mathbf{F}}_k\bar{\mathbf{B}}' - \tilde{\mathbf{F}}_k\tilde{\mathbf{B}}_g' \|_F^2,
\end{equation}
for $k=1,\ldots,2K$ and $g=1,2$ accordingly. Although there are alternative ways to refit the model from SCA, handling $\bar{\mathbf{F}}_k$ and $\tilde{\mathbf{F}}_k$ together has an advantage over treating them separately. First, \eqref{e:refitting} can be solved explicitly. Also, since the VAR in the factor models is latent, other refitting methods may be difficult to apply. Naturally, the squared error of the fitting becomes smaller after refitting the factor series by \eqref{e:refitting}. Then, we compute the empirical covariance matrix for the joint and group individual factor series, respectively. Finally, by the Yule-Walker equations \cite[e.g., Chapter 11 in][]{brockwell:2009}, we estimate the parameters in the model \eqref{e:filter_equation}, including the covariance matrix of noise $\bar{\boldsymbol{\Sigma}}_{\eta,k}$. We repeat this process for $k=1,\ldots,2K$ subjects.


\subsubsection{Estimation of loadings matrices and factor series}
\label{ssse:estimation_factor}

For the algorithm of estimating the joint loadings matrix and joint factor series at the second stage of our procedure, we use the direct fitting algorithm developed by \cite{kiers:1999}. Direct fitting means that one estimates components of factor-type models in the least squares sense, by regressing responses on covariates. It is different from the indirect fitting. \cite{harshman:1970} suggested an indirect fitting algorithm through alternating least squares (ALS) that can be described on the joint structure as follows. If we denote the covariance matrix of the joint component $\bar{\mathbf{X}}_k$ as $\bar{\boldsymbol{\Sigma}}_{k}$, the model can be estimated by minimizing
\begin{equation*}
    \min_{\bar{\mathbf{B}},\bar{\mathbf{C}}_{1:2K},\bar{\boldsymbol{\Phi}}} \sum_{k=1}^{2K}\|\bar{\boldsymbol{\Sigma}}_{k} - \bar{\mathbf{B}}\bar{\mathbf{C}}_k \bar{\boldsymbol{\Phi}}\bar{\mathbf{C}}_k\bar{\mathbf{B}}'\|_F^2.
\end{equation*}
Instead, we focus on the direct fitting algorithm. This algorithm has advantages because not only it has a fast convergence but it also produces the factor scores, which are used for constructing factor dynamics. Moreover, one can easily incorporate additional constraints such as non-negative entries in scaler matrices.

We now describe the direct fitting algorithm. From $\bar{\mathbf{F}}_{k} = \bar{\mathbf{A}}_{k}\bar{\mathbf{C}}_{k}$ where $\bar{\mathbf{A}}_{k}$ is a $T_k \times r_J$ matrix consisting of scaled factor series $\bar{F}_{t}^{(k)}$, the goal is to solve the following problem:
\begin{eqnarray}
 (\hat{\mathbf{A}}_{1:2K},\hat{\mathbf{C}}_{1:2K},\hat{\mathbf{B}}) 
 &=& \argmin_{\bar{\mathbf{A}}_{1:2K},\bar{\mathbf{C}}_{1:2K},\bar{\mathbf{B}}} \sum_{k=1}^{2K} \left\| \bar{\mathbf{X}}_k - \bar{\mathbf{A}}_k\bar{\mathbf{C}}_k\bar{\mathbf{B}}' \right\|_F^2, \label{e:PARAFAC2_orginal_object} \\
    &\textrm{s.t.}& \quad \bar{\mathbf{A}}_k'\bar{\mathbf{A}}_k = \bar{\mathbf{A}}_{k'}'\bar{\mathbf{A}}_{k'}, \quad k,k'\in\{1,\ldots,2K\}. \label{e:PARAFAC2_orginal_constraint}
\end{eqnarray}
The key idea is to convert the problem \eqref{e:PARAFAC2_orginal_object}--\eqref{e:PARAFAC2_orginal_constraint} into $2K$ separate orthogonal Procrustes problems, which have been dealt with in psychometrics \cite[e.g.,][]{green:1952,schonemann:1966}.

Note that the optimal solution \eqref{e:PARAFAC2_orginal_object}--\eqref{e:PARAFAC2_orginal_constraint} exists when $\bar{\mathbf{A}}_k'\bar{\mathbf{A}}_k = \bar{\mathbf{A}}'\bar{\mathbf{A}}$ for all $k$. Then, one can always find a columnwise orthonormal $T_k \times r_J$ matrix $\bar{\mathbf{P}}_k$ such that $\bar{\mathbf{A}}_k = \bar{\mathbf{P}}_k\bar{\mathbf{A}}$ \cite[e.g., Exercise 14.3.13 in][]{harville2006matrix} so that
\begin{equation*}
    \bar{\mathbf{A}}_k'\bar{\mathbf{A}}_k  = \bar{\mathbf{A}}'\bar{\mathbf{P}}_k'\bar{\mathbf{P}}_k\bar{\mathbf{A}} = \bar{\mathbf{A}}'\bar{\mathbf{A}}.
\end{equation*}
This corresponds to the necessary and sufficient conditions for the optimality described in Section 2 in \cite{kiers:1999}. Thus, the original problem \eqref{e:PARAFAC2_orginal_object}--\eqref{e:PARAFAC2_orginal_constraint} becomes
\begin{eqnarray}
    (\hat{\mathbf{P}}_{1:2K},\hat{\mathbf{A}},\hat{\mathbf{C}}_{1:2K},\hat{\mathbf{B}}) 
     &=& \argmin_{\bar{\mathbf{P}}_{1:2K},\bar{\mathbf{A}},\bar{\mathbf{C}}_{1:2K},\bar{\mathbf{B}}} \sum_{k=1}^{2K} \left\| \bar{\mathbf{X}}_k - \bar{\mathbf{P}}_k\bar{\mathbf{A}}\bar{\mathbf{C}}_k\bar{\mathbf{B}}' \right\|_F^2, \label{e:PARAFAC2_separate_object}\\
        &\textrm{s.t.}& \quad \bar{\mathbf{P}}_k'\bar{\mathbf{P}}_k = \mathbf{I}_{r_{J}},\quad k=1,\ldots,2K. \label{e:PARAFAC2_separate_constraint}
\end{eqnarray}
Note that the objective function for fixed $k$ is $\textrm{Tr}(\bar{\mathbf{X}}_k'\bar{\mathbf{X}}_k) - 2\textrm{Tr}(\bar{\mathbf{X}}_k'\bar{\mathbf{P}}_k\bar{\mathbf{A}}\bar{\mathbf{C}}_k\bar{\mathbf{B}}') + \textrm{Tr}(\bar{\mathbf{B}}\bar{\mathbf{C}}_k\bar{\mathbf{A}}'\bar{\mathbf{A}}\bar{\mathbf{C}}_k\bar{\mathbf{B}}')$, where $\textrm{Tr}(\cdot)$ is a trace operator. Hence, a minimizer of \eqref{e:PARAFAC2_separate_object}--\eqref{e:PARAFAC2_separate_constraint} over $\bar{\mathbf{P}}_k$ for fixed other components is obtained by
\begin{equation}\label{e:e:PARAFAC2_single}
    \hat{\mathbf{P}}_k = \argmax_{\bar{\mathbf{P}}_k'\bar{\mathbf{P}}_k = \mathbf{I}_{r_{J}}} \textrm{Tr}(\bar{\mathbf{A}}\bar{\mathbf{C}}_k\bar{\mathbf{B}}'\bar{\mathbf{X}}_k'\bar{\mathbf{P}}_k).
\end{equation}
Take the SVD as $\bar{\mathbf{A}}\bar{\mathbf{C}}_k\bar{\mathbf{B}}'\bar{\mathbf{X}}_k' = \bar{\mathbf{U}}_k\bar{\mathbf{S}}_k\bar{\mathbf{V}}_k'$, where $\bar{\mathbf{U}}_k$ and $\bar{\mathbf{V}}_k$ are $r_J \times r_J$ and $r_J \times T_k$ matrices with orthogonal columns and $\bar{\mathbf{S}}_{k}$ consists of singular values of $\bar{\mathbf{A}}\bar{\mathbf{C}}_k\bar{\mathbf{B}}'\bar{\mathbf{X}}_k'$. By replacing $\bar{\mathbf{A}}\bar{\mathbf{C}}_k\bar{\mathbf{B}}'\bar{\mathbf{X}}_k'$ with $\bar{\mathbf{U}}_k\bar{\mathbf{S}}_k\bar{\mathbf{V}}_k'$, the solution of \eqref{e:e:PARAFAC2_single} becomes $\bar{\mathbf{P}}_k = \bar{\mathbf{V}}_k\bar{\mathbf{U}}_k'$, which can be shown by using Cauchy–Schwarz inequality. This completes the first step of the direct fitting algorithm. Next, substituting $\bar{\mathbf{V}}_k'\bar{\mathbf{U}}_k$ for $\bar{\mathbf{P}}_k$ in \eqref{e:PARAFAC2_separate_object} leads to 
\begin{equation}\label{e:PARAFAC_object}
    (\hat{\mathbf{A}},\hat{\mathbf{C}}_{1:2K},\hat{\mathbf{B}}) 
    = \argmin_{\bar{\mathbf{A}},\bar{\mathbf{C}}_{1:2K},\bar{\mathbf{B}}} \sum_{k=1}^{2K} \left\| \hat{\mathbf{P}}_k'\bar{\mathbf{X}}_k - \bar{\mathbf{A}}\bar{\mathbf{C}}_k\bar{\mathbf{B}}' \right\|_F^2.
\end{equation}
The problem \eqref{e:PARAFAC_object} can be solved by CANDECOMP/PARAFAC (CP) model as the ALS algorithm \cite[CP-ALS, see][for the details]{kiers:1998}. This completes the second step. The two steps are repeated until a convergence criterion is satisfied. The same algorithm is applied to the estimates $\tilde{\mathbf{A}}_{k},\tilde{\mathbf{C}}_{k}$, and $\tilde{\mathbf{B}}_g$, with group individual components $\tilde{\mathbf{X}}_k$, $k=1,\ldots,2K$, for the corresponding $g=1,2$.

\subsubsection{Determinacy of loadings matrices and factor series}
\label{ssse:determinacy}

Determinacy of loadings and factors in SCA was studied by several researchers. \cite{harshman:1996} proved the uniqueness of a general form of parameters of multi-view data which led to the rotational determinacy of SCA-PF2 models. In particular, when the rank $r$ of factors exceeds 2, the estimated components multiplied with some scaling and permutation matrices described below can be determined if the number of subjects is greater than $r(r+1)(r+2)(r+3)/24$ for rank $r$. So, although the inequality seems to require a quite larger number of subjects to obtain the determinacy, the computational experiments by  \cite{ten:1996} suggested that the determinacy can be achieved even for larger ranks if the number of subjects exceeds 4. However, a rigorous proof for this or the other minimum number of subjects remains to be found.

The proved determinacy for SCA-PF2 is as follows. If the SCA-PF2 model can be expressed through two sets $\bar{\mathbf{B}},\bar{A}_t^{(k)},\bar{\mathbf{C}}_k$ and $\doublebar{\mathbf{B}},\doublebar{A}_t^{(k)},\doublebar{\mathbf{C}}_k$ of joint loadings matrices, scaled joint factor series, and the diagonal matrices, then 
\begin{equation}\label{e:rotational_determinacy_joint_1}
    \doublebar{\mathbf{B}} = \bar{\mathbf{B}}\bar{\mathbf{W}}\Delta_{\bar{\mathbf{B}}}, \quad 
    \doublebar{A}_t^{(k)} = z_k\Delta_{\bar{\mathbf{A}}}\bar{\mathbf{W}}'\bar{A}_t^{(k)},\quad 
    \doublebar{\mathbf{C}}_{k} = z_k\bar{\mathbf{W}}'\bar{\mathbf{C}}_k\bar{\mathbf{W}}\Delta_{\bar{\mathbf{C}}},
\end{equation}
where $\bar{\mathbf{W}}$ is a permutation matrix, $z_k=-1$ or $1$, and $\Delta_{\bar{\mathbf{A}}},\Delta_{\bar{\mathbf{B}}},\Delta_{\bar{\mathbf{C}}}$ are $r_{J} \times r_{J}$ non-singular diagonal scaling matrices satisfying $\Delta_{\bar{\mathbf{A}}}\Delta_{\bar{\mathbf{B}}}\Delta_{\bar{\mathbf{C}}}=\mathbf{I}_{r_{J}}$. By convention, one can fix $z_k=1$ for all $k$ to overcome the sign indeterminacy \cite[e.g.,][]{helwig:2013}. Thus, the joint factors can be related as
\begin{equation}\label{e:rotational_determinacy_joint_2}
    \doublebar{F}_t^{(k)} 
    := \doublebar{\mathbf{C}}_{k}\doublebar{A}_t^{(k)},
\end{equation}
and the correlation matrix of the scaled joint factors is of the form $\doublebar{\boldsymbol{\Phi}} = \Delta_{\bar{\mathbf{A}}}\bar{\mathbf{W}}' \bar{\boldsymbol{\Phi}} \bar{\mathbf{W}} \Delta_{\bar{\mathbf{A}}}$ so that 
\begin{equation}\label{e:rotational_determinacy_joint_3}
    \mathbb{E}\doublebar{F}_t^{(k)}\doublebar{F}_t^{(k)'}
    = \doublebar{\mathbf{C}}_{k}\doublebar{\boldsymbol{\Phi}}\doublebar{\mathbf{C}}_{k} 
    =  \underbrace{\bar{\mathbf{W}}'\bar{\mathbf{C}}_k\bar{\mathbf{W}}}_{=:\bar{\mathbf{C}}_k^{\bar{\mathbf{W}}}} \Delta_{\bar{\mathbf{B}}}^{-1} \underbrace{\bar{\mathbf{W}}'\bar{\boldsymbol{\Phi}}\bar{\mathbf{W}}}_{=:\bar{\boldsymbol{\Phi}}^{\bar{\mathbf{W}}}} \Delta_{\bar{\mathbf{B}}}^{-1} \bar{\mathbf{W}}'\bar{\mathbf{C}}_k\bar{\mathbf{W}} =\bar{\mathbf{C}}_k^{\bar{\mathbf{W}}} \Delta_{\bar{\mathbf{B}}}^{-1} \bar{\boldsymbol{\Phi}}^{\bar{\mathbf{W}}} \Delta_{\bar{\mathbf{B}}}^{-1} \bar{\mathbf{C}}_k^{\bar{\mathbf{W}}},
\end{equation}
where $\bar{\mathbf{C}}_k^{\bar{\mathbf{W}}}$ is again diagonal with permutated entries. That is, modulo the scaling matrix $\Delta_{\bar{\mathbf{B}}}$, one has the same covariance structure of the factor series up to their rearrangement. Note that from \eqref{e:rotational_determinacy_joint_1}--\eqref{e:rotational_determinacy_joint_3}, the determinacy can be achieved with the multiplication of the scalers and permutations. In the sense that although the model still has an issue of scaling, the identification up to the change of rows and columns of matrices by imposing SCA-PF2 structure is remarkable.

Similar results hold for group individual factor series. For each group $g=1,2$, the group loadings matrices $\tilde{\mathbf{B}}_g,\doubletilde{\mathbf{B}}_g$, scaled group individual factors $\tilde{A}_{t}^{(k)},\doubletilde{A}_{t}^{(k)}$, and the diagonal matrices $\tilde{\mathbf{C}}_k,\doubletilde{\mathbf{C}}_k$ are related as
\begin{equation*}
    \doubletilde{\mathbf{B}}_g = \tilde{\mathbf{B}}\tilde{\mathbf{W}}_g\Delta_{\tilde{\mathbf{B}}_g}, \quad 
    \doubletilde{A}_t^{(k)} = z_{g,k}\Delta_{\tilde{\mathbf{A}}_g}\tilde{\mathbf{W}}_g'\tilde{A}_t^{(k)},\quad 
    \doubletilde{\mathbf{C}}_{k} = z_{g,k}\tilde{\mathbf{W}}_g'\tilde{\mathbf{C}}_k\tilde{\mathbf{W}}_g\Delta_{\tilde{\mathbf{C}}_g}.
\end{equation*}
Here, $\tilde{\mathbf{W}}_g$ is a permutation matrix of $g$th group and $z_{g,k}=-1$ or $1$, so we take $z_{g,k}=1$ for all cases by convention. Obviously, the index $k$ goes from 1 to $K$ when $g=1$, and $k=K+1,\ldots,2K$ are used for $g=2$. Finally, $\Delta_{\tilde{\mathbf{A}}_g},\Delta_{\tilde{\mathbf{B}}_g},\Delta_{\tilde{\mathbf{C}}_g}$ are $r_{G} \times r_{G}$ non-singular diagonal matrices such that $\Delta_{\tilde{\mathbf{A}}_g}\Delta_{\tilde{\mathbf{B}}_g}\Delta_{\tilde{\mathbf{C}}_g}=\mathbf{I}_{r_{G}}$. Accordingly, the group individual factors are related as
\begin{equation*}
    \doubletilde{F}_t^{(k)} 
    = \doubletilde{\mathbf{C}}_{k}\doubletilde{A}_t^{(k)},
\end{equation*}
for corresponding $g$ and $k$ as above, so that the covariance of each group individual factor is 
\begin{equation*}
    \mathbb{E}\doubletilde{F}_t^{(k)}\doubletilde{F}_t^{(k)'}
    = \doubletilde{\mathbf{C}}_{k}\doubletilde{\boldsymbol{\Phi}}\doubletilde{\mathbf{C}}_{k} 
    =  \underbrace{\tilde{\mathbf{W}}_{g}'\tilde{\mathbf{C}}_k\tilde{\mathbf{W}_{g}}}_{:=\tilde{\mathbf{C}}_k^{\tilde{\mathbf{W}}_{g}}} \Delta_{\tilde{\mathbf{B}}}^{-1} \underbrace{\tilde{\mathbf{W}_{g}}'\tilde{\boldsymbol{\Phi}}\tilde{\mathbf{W}_{g}}}_{:=\tilde{\boldsymbol{\Phi}}^{\tilde{\mathbf{W}}_{g}}} \Delta_{\tilde{\mathbf{B}}}^{-1} \tilde{\mathbf{W}_{g}}'\tilde{\mathbf{C}}_k\tilde{\mathbf{W}_{g}} =\tilde{\mathbf{C}}_k^{\tilde{\mathbf{W}}_{g}} \Delta_{\tilde{\mathbf{B}}}^{-1} \tilde{\boldsymbol{\Phi}}^{\tilde{\mathbf{W}}_{g}} \Delta_{\tilde{\mathbf{B}}}^{-1} \tilde{\mathbf{C}}_k^{\tilde{\mathbf{W}}_{g}},
\end{equation*}
where $\doubletilde{\boldsymbol{\Phi}}_g := \Delta_{\tilde{\mathbf{A}}_g}\tilde{\mathbf{W}}_g' \tilde{\boldsymbol{\Phi}}_g \tilde{\mathbf{W}}_g \Delta_{\tilde{\mathbf{A}}_g}$, $g=1,2$. 

In Section \ref{se:illustrative} with simulation results, the determinacy conditions allow us to assess the estimation performance as follows. We start by permuting the columns of the estimated loadings matrices for each structure and determine the permutation orders by minimizing the mean squared errors to the true loadings. Subsequently, we rearrange the factor series in each structure based on these orders.

\subsection{Initial rank selection}
\label{ssse:initial_rank}

As seen in \eqref{e:mat_J}, AJIVE requires an initial guess $\hat{r}_k$ of the rank $r_k$. This is required not only for AJIVE but also for other estimation algorithms of reduced-rank models. The rank can be estimated through various methods. We acknowledge that as the number of subjects increases, deciding on the initial rank using graphical tools like a scree plot becomes more challenging. Moreover, even when employing cross-validation-based rank selection algorithms \cite[e.g.,][]{bro:2003}, our model encompasses several distinct structures within the joint and two-group individual structures, with all structures being interrelated. Furthermore, our application involves a large number of subjects, making the implementation of cross-validation complex.

Instead, we opt for the rotational bootstrap algorithm developed by \cite{prothero:2022}. While this algorithm was originally designed to identify partially shared structures, it can also be used to explore the best reduced-rank approximation of the true block signals. This algorithm is also rooted in PAA, and involves bootstrap to construct an upper bound for the largest principal angle between the subspace of the unknown signal block and the low-rank approximation of the observation block. We apply the algorithm to each block observation and simplify notation by omitting the subscript $k$ for the subject number.

The first step of the algorithm is to compute shrunk singular values and imputed noise matrices. Suppose that a data block $\mathbf{X}=\mathbf{Y}+\mathbf{E}$ is given. Consider the SVD $\hat{\mathbf{X}} = \hat{\mathbf{U}}\hat{\mathbf{S}}\hat{\mathbf{V}}'$, where $\hat{\mathbf{U}},\hat{\mathbf{V}}$ consist of orthonormal columns $\hat{U}_i$ and $\hat{V}_i$, and diagonal matrix $\hat{\mathbf{S}}$ with singular values $\sigma_1 \geq \ldots \geq \sigma_{d\wedge T}$, where $d\wedge T$ is the minimum between the number of variables $d$ and the sample length $T$ for the subject of interest. Then, the thresholded singular values are $\check{\sigma}_i := \check{\kappa}h^{*}(\sigma_{i}/\check{\kappa})$, $i=1,\ldots,d \wedge T$, where $\check{\kappa}=\frac{\sigma_{\textrm{med}}}{\sqrt{\textrm{MP}(\beta)_{0.5}}}$ for $\sigma_{\textrm{med}} = \textrm{med}(\sigma_1,\ldots,\sigma_{d\wedge T})$, $\textrm{MP}(\beta)_{q}$ is the 100$q$ percentile (hence, the median for $\check{\kappa}$) of the Marchenko-Pastur distribution with the so-called aspect ratio $\beta=\frac{d \wedge T}{d \vee T}$ \cite[e.g.,][]{gavish:2014}, and the thresholding function used is 
\begin{equation*}
    h^{*}(a) = \left\{\begin{array}{ll}
      \frac{1}{\sqrt{2}}\sqrt{a^2-\beta-1+\sqrt{(a^2-\beta-1)^2-4\beta}} ,  & \textrm{if }a\geq 1+\sqrt{\beta}, \\
      0,   & \textrm{if } a<1+\sqrt{\beta}.
    \end{array}\right.
\end{equation*}
Naturally, the maximal allowable rank $\check{r}$ is determined by the number of non-zero $\check{\sigma}_i$'s. Thus, $\check{\mathbf{S}}_{\check{r}}$ represents a $\check{r}$-dimensional diagonal matrix with entries $\check{\sigma}_r$ for $r=1,\ldots,\check{r}$. The estimate $\check{\mathbf{E}}$ of the noise matrix $\mathbf{E}$ for bootstrapping is computed by 
\begin{equation*}
    \check{\mathbf{E}} 
    = \sum_{i=1}^{\check{r}}\check{\kappa}\textrm{MP}(\beta)_{u_i}\hat{U}_{i}\hat{V}_{i}' + \sum_{i=\check{r}+1}^{d\wedge T}\sigma_i\hat{U}_{i}\hat{V}_{i}',
\end{equation*}
where $u_{1},\ldots,u_{\check{r}}$ are i.i.d. uniform random variables on $(0,1)$. This estimate is thought of as imputed noise, which is justified in Appendix C in \cite{prothero:2022}.

The rest of the algorithm consists of two nested loops. In the outer loop, obtain $T \times \check{r}$ and $d \times \check{r}$ independently generated orthonormal matrices, $\bar{\mathbf{U}}_{\check{r}}^{(\ell)}$ and $\bar{\mathbf{V}}_{\check{r}}^{(\ell)}$, $\ell=1,\ldots,L$, and construct $\bar{\mathbf{X}}^{(\ell)} = \bar{\mathbf{U}}_{\check{r}}^{(\ell)}\check{\mathbf{S}}_{\check{r}}\bar{\mathbf{V}}_{\check{r}}^{(\ell)'}+\check{\mathbf{E}}$. Then, compute matrices with orthonormal columns associated with $r$ largest singular values from the SVD of $\bar{\mathbf{X}}^{(\ell)}$, denoted as $\hat{\mathbf{U}}_{r}^{(\ell)}$ and $\hat{\mathbf{V}}_{r}^{(\ell)}$, $r=1,\ldots,\check{r}$. In the inner loop, compute the smallest singular values $\sigma_{\min}(\cdot)$ from SVDs of $\bar{\mathbf{U}}_{\check{r}}^{(\ell)'}\hat{\mathbf{U}}_{r}^{(\ell)}$ and $\bar{\mathbf{V}}_{\check{r}}^{(\ell)'}\hat{\mathbf{V}}_{r}^{(\ell)}$ for $r=1,\ldots,\check{r}$ and then, convert them into the largest principal angles $\theta_{U,r}^{(\ell)}$ and $\theta_{V,r}^{(\ell)}$ between $\bar{\mathbf{U}}_{\check{r}}^{(\ell)}$ and $\hat{\mathbf{U}}_{r}^{(\ell)}$, and $\bar{\mathbf{V}}_{\check{r}}^{(\ell)}$ and $\hat{\mathbf{V}}_{r}^{(\ell)}$ by $\arccos(\sigma_{\min}(\bar{\mathbf{U}}_{\check{r}}^{(\ell)'}\hat{\mathbf{U}}_{r}^{(\ell)}) )$ and $\arccos(\sigma_{\min}(\bar{\mathbf{V}}_{\check{r}}^{(\ell)'}\hat{\mathbf{V}}_{r}^{(\ell)}) )$, respectively. This leads to $L \times \check{r}$ computed principal angles $\theta_{U,r}^{(\ell)}$ and $\theta_{V,r}^{(\ell)}$ for each of the row and column spaces. After sorting those angles for each $r$, select the final initial rank estimate $\hat{r}$ by 
\begin{equation*}
    \hat{r} 
    = \min \left( 
    \sum_{r=1}^{\check{r}}
    1_{\{ [ \textrm{95 percentile of } \theta_{U,r}^{(\ell)}, \ \ell=1,\ldots,L] \ < \ \xi\theta_{0,U} \} } , \ 
    \sum_{r=1}^{\check{r}} 1_{\{ [ \textrm{95 percentile of } \theta_{V,r}^{(\ell)}, \ \ell=1,\ldots,L] \ < \ \xi\theta_{0,V} \} } 
    \right).
\end{equation*}
Here, $\theta_{0,U}$ and $\theta_{0,V}$ represent the random direction angle bounds for the row and column subspaces and are computed in parallel in the inner loop of the bootstrapping algorithm. Similarly to AJIVE, the candidates of the random direction bounds are calculated as the largest principal angles between $\vec{\mathbf{U}}_{\check{r}}^{(m)}$ and $\vec{\mathbf{U}}_{r}^{(m)}$, and between $\vec{\mathbf{V}}_{\check{r}}^{(m)}$ and $\vec{\mathbf{V}}_{r}^{(m)}$, where $\vec{\mathbf{U}}_{r}^{(m)}$ and $\vec{\mathbf{V}}_{r}^{(m)}$ are independently generated orthonormal matrices of the sizes $T \times r$ and $d \times r$, $r=1,\ldots,\check{r}$, respectively. The algorithm repeats this computation $m=1,\ldots,M(=L)$ times and selects the 5 percentile of all replications. The tuning parameter $\xi$, when multiplied by the bounds, controls the maximum allowed principal angles. An increase in the value of $\xi$ leads to larger angles falling within the criterion, typically resulting in higher estimated initial ranks. The method for fine-tuning $\xi$ may vary depending on the specific case and remains somewhat unclear. However, it is generally recommended to keep it within the range of $(0, 0.5]$. In our empirical results in Section \ref{sse:result_bootstrap} and the application illustrated in Section \ref{se:data_app}, we set $\xi = 0.5$, which has shown reasonable performance. We refer to Algorithm 1 in Section 2.1.3 in \cite{prothero:2022} for the remaining details.

In practice, the rotational bootstrap is applied to the block observation of each subject. Then, the initial rank is finalized by a majority vote, as a mode of $2K$ number of $\hat{r}$'s. We use this estimate for selecting the rank of the joint factors $\hat{r}_J$ produced by AJIVE. Naturally, this leads to $\hat{r}_G=\hat{r}-\hat{r}_J$. Potentially, one can also take different initial ranks for each group, or even different initial ranks for each subject, depending on the context. As noted in Section \ref{ssse:estimation_structure}, while AJIVE does not require identical initial ranks across subjects and neither does the rotational bootstrap, we assume that the initial ranks are identical at least within each group. In order to apply the factor structure described below for the rest of the procedure, the ranks of the joint and the two group blocks should be specified. Furthermore, to impose determinacy conditions across subjects which are described below, it is convenient for the ranks within each group to be identical in that SCA described below can be employed for both joint and group individual structures. The identification of factor structure relies on the performance of the initial rank selection from the rotational bootstrap and the joint rank selection from AJIVE. We will assess the performance of the initial rank selection in Section \ref{sse:result_bootstrap}.


\section{Illustrative examples}
\label{se:illustrative}

\subsection{Simulation design}
\label{sse:design}

In this section, we describe the simulation design that is used to assess estimation performance under different scenarios. The results are reported in Sections \ref{sse:result_main} and \ref{sse:result_bootstrap} below. The simulation study has several settings. First, we vary $d$, $T$, and $K$ related to the size of the problem to see whether the estimators exhibit what one can expect from the standard factor models. Second, for the fixed size of the problem, inspired by the group-level analysis, we vary the signal-to-noise ratio (SNR) to check the effect on model recovery. Third, we examine how well the proposed bootstrap procedure identifies the initial ranks. The R package \texttt{multiway} and other supplementary code of \cite{helwig:2019}, and the R package \texttt{ajive} of \cite{Carmichael:2020} were used in simulations. For \texttt{ajive}, since the last maintenance was many years ago, we extracted and reorganized the functions for the estimation.

The data generating processes (DGPs) are as follows.
\begin{itemize}
    \item[-] $X_{i,t}^{(k)} 
    = \left\{\begin{array}{ll}
    \sum_{j=1}^{r_J}\bar{B}_{i,j}\bar{F}_{j,t}^{(k)} + \sum_{j=1}^{r_G}\tilde{B}_{i,j}^{(1)}\tilde{F}_{j,t}^{(k)} + E_{i,t}^{(k)}, & \textrm{if }k=1,\ldots,K, \\
    \sum_{j=1}^{r_J}\bar{B}_{i,j}\bar{F}_{j,t}^{(k)} + \sum_{j=1}^{r_G}\tilde{B}_{i,j}^{(2)}\tilde{F}_{j,t}^{(k)} + E_{i,t}^{(k)}, & \textrm{if }k=K+1,\ldots,2K,
    \end{array}\right.$
    
    \item[-] Among the indices $i=1,\ldots,d$, choose 50\% at random and make $\bar{B}_{i,j}$, $j=1,\ldots,r_J$, non-zero as $\bar{B}_{i,j}\stackrel{i.i.d.}{\sim}\mathrm{Unif}(0,1)$. Otherwise, $\bar{B}_{i,j}=0$ for the non-selected indices $i$. This ensures identifiability as discussed in Section \ref{ssse:identifiability}.
    
    \item[-] Among the non-selected indices $i$ above, half of them are chosen to make $\tilde{B}_{i,j}^{(1)}$, $j=1,\ldots,r_G$, have non-zero entries. The remaining half are used for $\tilde{B}_{i,j}^{(2)}$, $j=1,\ldots,r_G$, to be non-zero. Both are generated by $\tilde{B}_{i,j}^{(1)},\tilde{B}_{i,j}^{(2)}\stackrel{i.i.d.}{\sim}\mathrm{Unif}(0,1)$. Otherwise, let $\tilde{B}_{i,j}^{(1)},\tilde{B}_{i,j}^{(2)}$ be zero for the non-selected indices of $i$.
    
    \item[-] For each $k=1,\ldots,2K$, generate the series $\bar{A}_{t}^{(k)}$ and $\tilde{A}_{t}^{(k)}$, $t=1,\ldots,T_{k}$, from VAR(1) models,
    \begin{eqnarray*}
        \bar{A}_{t}^{(k)} 
        = \bar{\boldsymbol{\Psi}}\bar{A}_{t-1}^{(k)} + \bar{\xi}_{t}^{(k)},\quad \{\bar{\xi}_{t}^{(k)}\} \sim \mathrm{WN}(0,\sigma_{\bar{\xi}}\mathbf{I}_{r_{J}}), \\
        \tilde{A}_{t}^{(k)} 
        = \tilde{\boldsymbol{\Psi}}\tilde{A}_{t-1}^{(k)} + \tilde{\xi}_{t}^{(k)},\quad \{\tilde{\xi}_{t}^{(k)}\} \sim \mathrm{WN}(0,\sigma_{\tilde{\xi}}\mathbf{I}_{r_{G}}).        
    \end{eqnarray*}
    For simplicity, let $\sigma_{\bar{\xi}} = \sigma_{\tilde{\xi}} = \sigma_{\xi}$. Let also assume $\bar{\boldsymbol{\Psi}}$ and $\bar{\boldsymbol{\Phi}}=\mathbb{E}\bar{A}_t^{(k)}\bar{A}_t^{(k)'}$ have the same structures as $\tilde{\boldsymbol{\Psi}}$ and $\tilde{\boldsymbol{\Phi}}=\mathbb{E}\tilde{A}_t^{(k)}\tilde{A}_t^{(k)'}$, respectively. For stable factor series, the VAR transition matrices $\bar{\boldsymbol{\Psi}}$ and $\tilde{\boldsymbol{\Psi}}$ satisfy the following equations,
    \begin{eqnarray}
        \bar{\boldsymbol{\Phi}} 
        &=& \bar{\boldsymbol{\Psi}} \bar{\boldsymbol{\Phi}} \bar{\boldsymbol{\Psi}}^{'} + \sigma_{\xi}\mathbf{I}_{r_{J}}, \label{e:riccati_joint} \\
        \tilde{\boldsymbol{\Phi}} 
        &=& \tilde{\boldsymbol{\Psi}} \tilde{\boldsymbol{\Phi}} \tilde{\boldsymbol{\Psi}}^{'} + \sigma_{\xi}\mathbf{I}_{r_{G}}. \label{e:riccati_group}
    \end{eqnarray}
    From the known fact of algebraic Riccati equations, the stable solutions $\bar{\boldsymbol{\Psi}}$ and $\tilde{\boldsymbol{\Psi}}$ can be uniquely determined \cite[e.g.,][]{boyd:1994}. To ensure the stability of the factor series, we fix $\bar{\boldsymbol{\Phi}}$ and $\tilde{\boldsymbol{\Phi}}$ described below first and solve the equation \eqref{e:riccati_joint}--\eqref{e:riccati_group} to obtain the VAR transition matrices. Different structures of $\bar{\boldsymbol{\Phi}}$ and $\tilde{\boldsymbol{\Phi}}$ will be referred to as types below.
    
    \item[-] Take $\bar{F}_{t}^{(k)} = \bar{\mathbf{C}}_k\bar{A}_{t}^{(k)}$ and $\tilde{F}_{t}^{(k)} = \tilde{\mathbf{C}}_k\tilde{A}_{t}^{(k)}$, where $\bar{\mathbf{C}}_k$ and $\tilde{\mathbf{C}}_k$ are diagonal and
    \begin{equation*}
        \bar{\mathbf{C}}_k = \tilde{\mathbf{C}}_k = \sqrt{c}\ \textrm{diag}((4+1),\ldots,(4+r)),
    \end{equation*}
    where $c>0$ is a SNR parameter, $r$ is the number of factor series, $r=r_{J}$ for $\bar{A}_{t}^{(k)}$ and $r=r_G$ for $\tilde{A}_{t}^{(k)}$.
    
    \item[-] $E_{t}^{(k)}\sim\mathcal{N}\left(0,\Sigma_{\mathbf{E},k} = \sigma_{\varepsilon} \mathbf{I}_{d} \right)$ with $\sigma_{\varepsilon}=1$, $k=1,\ldots,2K$.
    
\end{itemize}
Note that the necessity condition of the model identifiability in Section \ref{ssse:identifiability} is satisfied by having non-overlapping non-zero rows in loadings matrices. In summary, DGP is fully controlled by $(\textrm{type},c,\sigma_{\xi})$ for fixed $(d,T,K,r_J,r_G)$. By changing the size of the problem and the parameters, we vary the generated series to see how the realizations with different characteristics affect estimation performance. 

We assess estimation performance in three ways involving 5 measures of interest. First, we report $R^2$ statistics to measure how much each structure contributes to explain the variability of $\mathbf{X}_k$. For the joint structure, define this measure by
\begin{equation*}
    R^2 = \frac{1}{2K}\sum_{k=1}^{2K}\left( 1-\frac{\|\mathbf{X}_k - \hat{\bar{\mathbf{F}}}_k\hat{\mathbf{B}}\|_F^2}{\|\mathbf{X}_k\|_F^2} \right),
\end{equation*}
where $\hat{\bar{\mathbf{F}}}_k$ is the output from either PARAFAC2 fitting \eqref{e:PARAFAC2_orginal_object}--\eqref{e:PARAFAC2_orginal_constraint}, the subject-wise linear regression \eqref{e:refitting}, or the benchmark methods. Replacing $\hat{\bar{\mathbf{F}}}_k\hat{\bar{\mathbf{B}}}^{'}$ and $2K$ by $\hat{\tilde{\mathbf{F}}}_k\hat{\tilde{\mathbf{B}}}_g^{'}$ and $K$, gives the $R^2$ statistics of the group individual structures for $g=1,2$.

Another measure of the model fit is a root mean squared error (RMSE). RMSE considers the fitness of both the joint and group individual components to their model counterparts, rather than the component variability as in $R^2$ statistics. For the joint structure, we define
\begin{equation*}
    RMSE = \sqrt{\frac{1}{2dK}\sum_{k=1}^{2K}\frac{1}{T_k}\| \bar{\mathbf{X}}_k - \hat{\bar{\mathbf{F}}}_k\hat{\bar{\mathbf{B}}}^{'}\|_F^2}.
\end{equation*}
Replacing $\bar{\mathbf{X}}_k$, $\hat{\bar{\mathbf{F}}}_k\hat{\bar{\mathbf{B}}}^{'}$ and $2K$ by $\tilde{\mathbf{X}}_k$, $\hat{\tilde{\mathbf{F}}}_k\hat{\tilde{\mathbf{B}}}_g^{'}$ and $K$, gives the RMSE of the group individual structures for  $g=1,2$.

Finally, at each component level, we compute three measures of how estimators capture spatial information and temporal dependence. We introduce the Tucker congruence coefficient $(CC_M)$ defined as
\begin{equation*}
    CC_M = \frac{\mathbf{M}'\hat{\mathbf{M}}}{\sqrt{\mathbf{M}'\mathbf{M}}\sqrt{\hat{\mathbf{M}}'\hat{\mathbf{M}}}},
\end{equation*}
where $\mathbf{M}$ is a parameter vector and $\hat{\mathbf{M}}$ corresponds to its estimate. We consider loadings matrices, factor series, and VAR transition matrices in place of $\mathbf{M}$, with their corresponding performance measures denoted by $CC_B$, $CC_{F}$, and $CC_{\Psi}$, respectively. For loadings matrices, the congruence coefficients are computed for each column, and their mean is reported. Analogous calculations are performed on the columns of the factor series, but only the average over all subjects is reported. Note that $CC_{F}$ may inherently vary more, as the measure compares two random processes. Nevertheless, we expect that these $CC_F$ scores are still informative. For VAR transition matrices, we vectorize them to compute $CC_{\Psi}$. The CC scores will be reported for the joint and two groups separately.

For each simulation setting, we replicate the data generation and estimation 100 times. For the first simulation setting, we fix $c=1$ and consider $\textrm{type}=1$ and 2, where the types are described below. For the size of the problem, we let $(d,T,K)$ be:
\begin{itemize}
    \item Growing numbers of variables: Fix $T=200$, $K=50$. Take $d=200$ and 400.
    
    \item Growing observation times: Fix $d=100$, $K=50$. Take $T=200$ and $400$.
    
    \item Growing numbers of subjects: Fix $d=100$, $T=200$. Take $K=10$ and $100$.
\end{itemize}
For each case, we fix the ranks as $(r_J, r_G) = (2, 2)$ so that, as we can see from the results in Section \ref{sse:result_bootstrap}, the sums of the given ranks are guaranteed to be found through the initial rank selection algorithm.

For the second setting, we define the SNR by
\begin{equation}
    \textrm{SNR} = \frac{\|\bar{\mathbf{B}}\bar{\mathbf{C}}_k \bar{\boldsymbol{\Phi}} \bar{\mathbf{C}}_k\bar{\mathbf{B}}' + \tilde{\mathbf{B}}_g\tilde{\mathbf{C}}_k \tilde{\boldsymbol{\Phi}} \tilde{\mathbf{C}}_k\tilde{\mathbf{B}}_g'\|_F}{\|\Sigma_{\mathbf{E},k}\|_F} = f(c),
\end{equation}
as a function of $c=0.25,0.5,0.75,1.25,2$, and $4$ in our simulation. The type refers to the correlation structure $\bar{\boldsymbol{\Phi}}$ and $\tilde{\boldsymbol{\Phi}}$ between the factor series in each structure. $\bar{\boldsymbol{\Phi}}$ and $\tilde{\boldsymbol{\Phi}}$ for type 1 are
\begin{equation}\label{e:covariance_example}
    \bar{\boldsymbol{\Phi}},\tilde{\boldsymbol{\Phi}} 
    = 1,\quad 
    \begin{pmatrix}
    1 & -0.6 \\
    -0.6 & 1
    \end{pmatrix},\quad
    \begin{pmatrix}
    1 & -0.6 & 0.3 \\
    -0.6 & 1 & -0.6 \\
    0.3 & -0.6 & 1
    \end{pmatrix},\quad
    \begin{pmatrix}
    1 & -0.6 & 0.3 & -0.1\\
    -0.6 & 1 & -0.6 & 0.3\\
    0.3 & -0.6 & 1 & -0.6\\
    -0.1 & 0.3 & -0.6 & 1
    \end{pmatrix},
\end{equation}
with $\sigma_{\xi}=0.2$ and the correlation structures for type 2 are the identity matrices with the corresponding ranks, where $\sigma_{\xi}=0.3$.

We consider six other methods as benchmarks for comparison. The first group of alternative methods consists of replacing the PARAFAC2 fitting for joint and group individual components at the second stage of GRIDY with another procedure. Here, we use \textit{SCA-P} and \textit{GICA}. Note that SCA-P fitting follows a similar algorithm as PARAFAC2 fitting described in Section \ref{ssse:estimation_factor} but with simplified steps since the constraint \eqref{e:PARAFAC2_orginal_constraint} need not be considered. Once the estimates are produced, they are refitted by using varimax rotation of \cite{kaiser:1958}. For fitting GICA, we follow the estimation through the tensor decomposition of \cite{beckmann:2005}. The R code is provided in the supplementary material of \cite{helwig:2019} and the detailed discussion of the model is included in \cite{helwig:2013}. Another benchmarks are to use the SCA-PF2 and GICA model at the first two stages of our approach. Consider, for example, PARAFAC2 fitting. Recall that the first two stages consist of fitting PARAFAC2 models to three factorized data blocks (joint component block and two group individual component blocks) obtained by AJIVE. However, motivated by practical considerations, one could fit PARAFAC2 to the whole data block first. The extracted signal block naturally becomes the joint structure. Then, for the remainder of each group (that is, the difference between the data and the joint block), one could fit PARAFAC2 again to obtain the group individual structure. The rest of the steps are the same as GRIDY. The same approach is possible with the GICA model as well. We refer to these benchmarks as \textit{double-SCA} (\textit{DSCA}) and \textit{double-GICA} (\textit{DGICA}), respectively. However, we stress that these types of approaches have an issue, despite being used by researchers, since in order to fit PARAFAC2 models, one has to know the ranks for the factors in advance. We note that the AJIVE implicitly selects the joint rank and this joint rank estimation is quite reliable as shown in the literature. For this comparison, we assume that the joint rank is explicitly known.

Finally, one can directly use the estimated factor series from PARAFAC2 or GICA fitting at the second stage of our approach. That is, one can omit the refitting of factor series in \eqref{e:refitting}. We denote these benchmarks as \textit{Unfitted\_SCA-PF2} and \textit{Unfitted\_GICA}, respectively. In summary, the first two benchmarks focus on different factor models. The next two benchmarks concern cases where other decomposition methods are used, while the models of the factors remain the same. The last two benchmarks focus on the absence of a refitting step for the same factor models. The similarities and differences of the estimation procedures are summarized in Table \ref{tab:summary} of the supplemental material.

\subsection{Simulation results}
\label{sse:result_main}

We present the simulation results in the order of the settings described in Section \ref{sse:design}. All figures are displayed at the end of this work. Each of the figures consists of 5 rows, each of which represents $R^2$ statistics, RMSEs, and three Tucker congruence coefficients for the loadings $(CC_B)$, factor series $(CC_F)$, and VAR transition matrices $(CC_{\Psi})$. Columns in each figure present the measures for joint, group individual 1, and group individual 2 structures, respectively.

\begin{figure}[]
     \centering
     \includegraphics[width=1\textwidth,height=0.8\textheight]{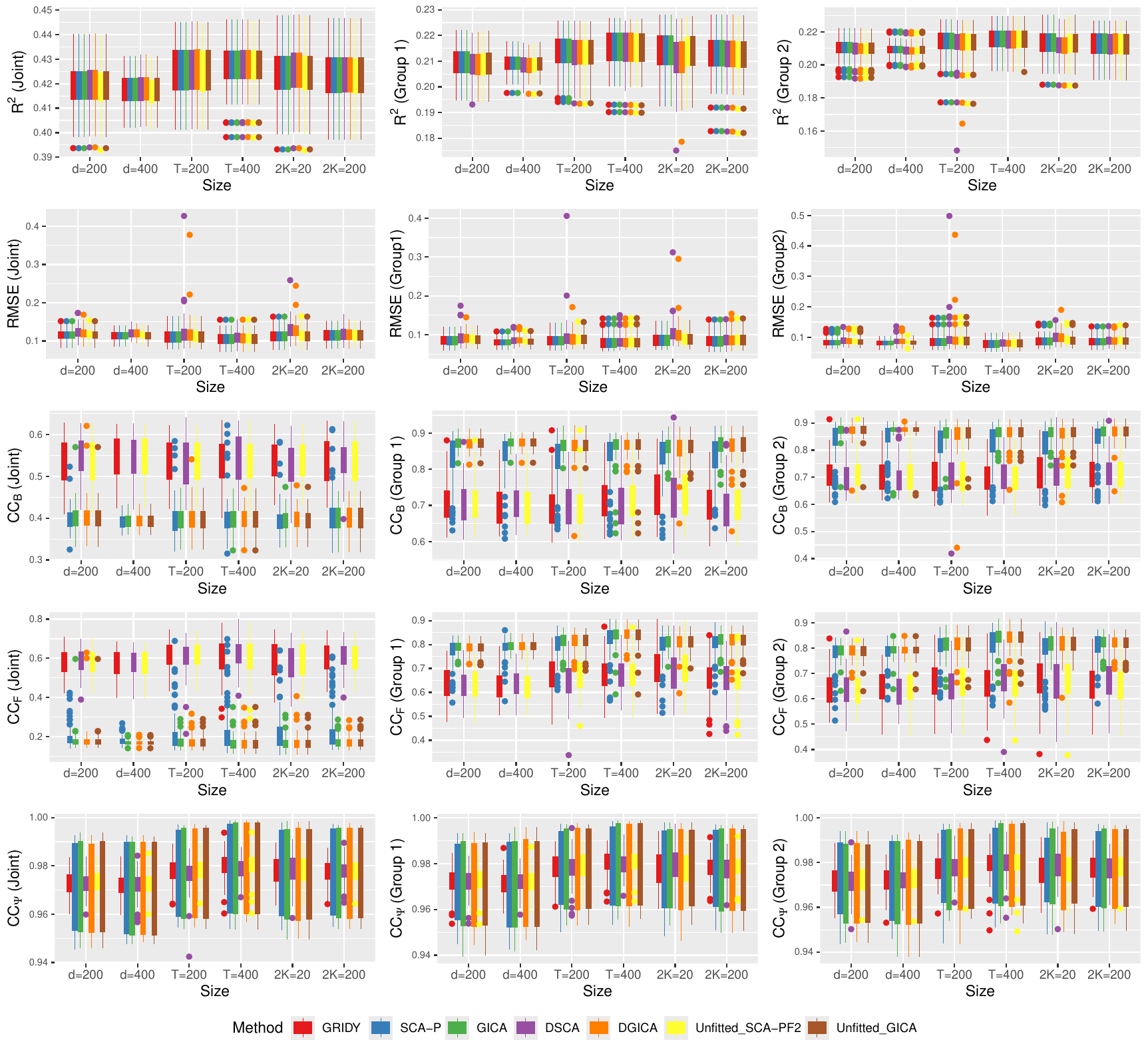}
     \caption{Box plots of the first result of the first simulation setting as one of the numbers of variables $d$, the numbers of samples $T$, and the numbers of subjects $K$ increases while the other two are fixed. The correlation type is 1, the ranks used in the simulation are $(r_J,r_G)=(2,2)$, and the SNR parameter is set to be $c=1$. Details of the rest of the model parameters are described in Section \ref{sse:design}. The leftmost red box for each figure represents GRIDY, while the other six boxes with abbreviations and different colors in the legend represent benchmarks.}
      \label{fig:Figure1}
\end{figure}

Figure \ref{fig:Figure1} presents the first simulation setting where correlations between the factor series within each structure are allowed. Regarding the sizes of problems, either the number of variables $d$, the number of samples $T$, or the number of subjects $K$ varies while the rest among $d,T$, and $K$ are fixed. As $d$ and $T$ increase, the variabilities of $R^2$ statistics and RMSEs tend to decrease, whereas they increase as $K$ increases. Having more subjects thus makes the analysis more challenging. Another interesting point is that DSCA and DGICA tend to have slightly higher $R^2$ statistics and RMSEs for the joint structures. For the group individual structures, on the other hand, both $R^2$ statistics and RMSEs are worse compared to any other benchmarks. Next, for the three congruence coefficients, GRIDY, DSCA, and Unfitted\_SCA-PF2 seem to recover the joint loadings matrices and joint factor series better compared to other benchmarks, while they are less capable of identifying group individual loadings and group individual factor series. We conjecture that this phenomenon is related to the estimation procedure, in that the three methods, including GRIDY, use PARAFAC2 fitting, while the rest rely on SCA-P fitting or its variants. Lastly, most of the methods can identify the VAR transition matrix of the factor series. However, GRIDY and DSCA show less variability in the measure.

\begin{figure}[]
     \centering
     \includegraphics[width=1\textwidth,height=0.8\textheight]{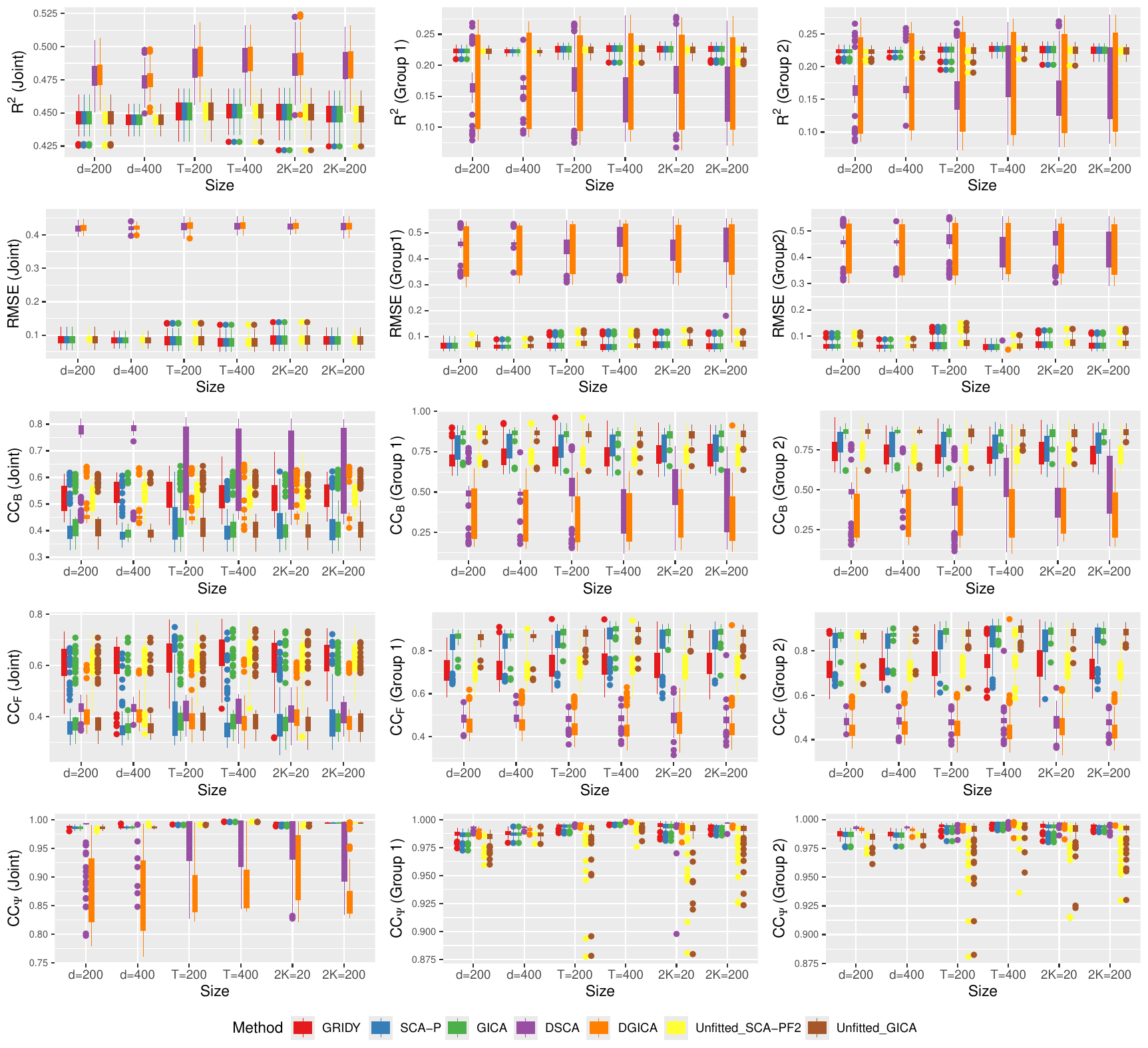}
     \caption{Box plots of the second result of the first simulation setting as one of the numbers of variables $d$, the numbers of samples $T$, and the numbers of subjects $K$ increases while the other two are fixed. The correlation type is 2, the ranks used in the simulation are $(r_J,r_G)=(2,2)$, and the SNR parameter is set to be $c=1$. Details of the rest of the model parameters are described in Section \ref{sse:design}. The leftmost red box for each figure represents GRIDY, while the other six boxes with abbreviations and different colors in the legend represent benchmarks.}
      \label{fig:Figure2}
\end{figure}

Figure \ref{fig:Figure2} depicts the simulation results for the same setting as Figure \ref{fig:Figure1}, but assuming that the factor series within structures are uncorrelated. Interestingly, the landscapes of performance measures from the two figures look quite different. When there are no correlations between the factors (Figure \ref{fig:Figure2}), the differences between the methods become more pronounced. The current simulation setting with the rank $r=r_J+r_G=4$ is typical to factor model literature. However, as we can see in the subfigures for $R^2$ statistics and RMSEs, the values for increasing $d$, $T$, or $K$ vary little, unlike what we would expect in standard factor models \cite[e.g., Table 1 in][]{doz:2011}. Among the methods, DSCA and DGICA seem to struggle to fit the model, while the rest of the methods tend to perform similarly as in the previous figure. In data applications, where we cannot assume the correlation structure of the factor series in advance, it would be beneficial to use GRIDY or other benchmarks that use a similar estimation procedure as GRIDY. Additionally, it seems advantageous to refit the estimates from PARAFAC2 or GICA when identifying the VAR transition matrices of the factor series, as seen by the worse performance for $CC_{\Psi}$ in the unfitted cases. Finally, among GRIDY, SCA-P, and GICA, while the other two show the larger variability in this type of correlation structure, our proposed method is the most robust in the sense that the performance is similar across the different types. 

\begin{figure}[]
     \centering
     \includegraphics[width=1\textwidth,height=0.8\textheight]{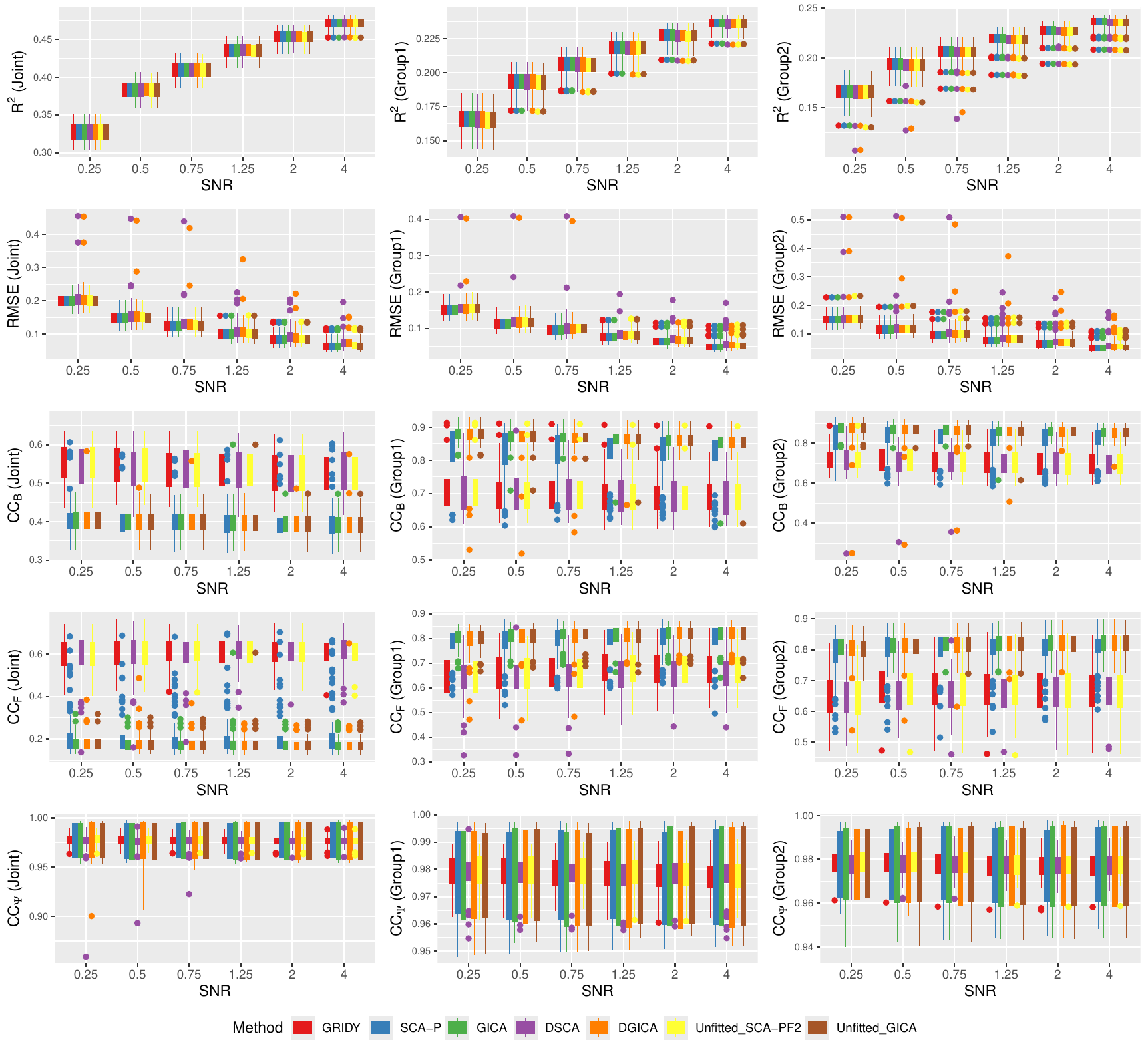}
     \caption{Box plots of the first result of the second simulation setting as SNR parameter $c$ increases from $c=0.25$ to $c=4$. The correlation type is 1, the size of problems is fixed by $(d,T,K)=(100,200,50)$, and the ranks used in the simulation are $(r_J,r_G)=(2,2)$, and the SNR is set to be $c=1$. Details of the model parameter are described in Section \ref{sse:design}. The leftmost red box for each figure represents GRIDY, while the other six boxes with abbreviations and different colors in the legend represent benchmarks.}
      \label{fig:Figure3}
\end{figure}

The next two figures concern the second simulation setting. Figure \ref{fig:Figure3} presents the results for SNR parameter $c$ increasing from $c=0.25$ to $c=4$. Notably, unlike in the previous two simulation results, $R^2$ statistics and RMSEs tend to improve along the increase of the SNR parameter. We also see that their variabilities are decreasing. However, the congruence coefficients for the loadings matrices, factor series, and VAR transition matrices seem to be little influenced by the increase in the strength of the signals. This might be related to the fact that the fitting algorithm factorizes the estimated blocks, not the (smoothed) raw block observations \cite[e.g., see Figure 3 in][]{helwig:2019}.

\begin{figure}[]
     \centering
     \includegraphics[width=1\textwidth,height=0.8\textheight]{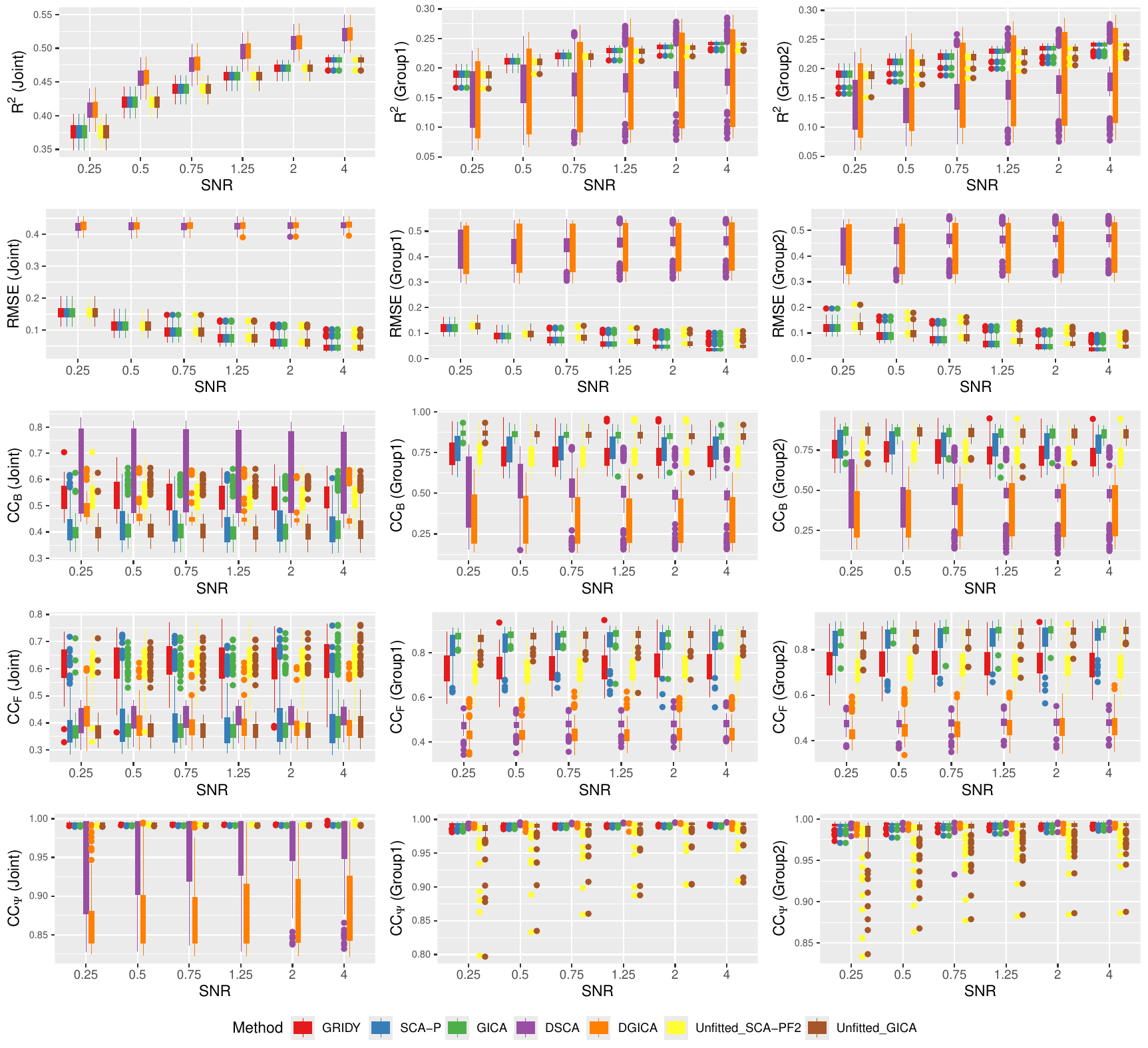}
     \caption{Box plots of the first result of the second simulation setting as SNR parameter $c$ increases from $c=0.25$ to $c=4$. The correlation type is 2, the size of problems is fixed by $(d,T,K)=(100,200,50)$, and the ranks used in the simulation are $(r_J,r_G)=(2,2)$, and the SNR is set to be $c=1$. Details of the model parameter are described in Section \ref{sse:design}. The leftmost red box for each figure represents GRIDY, while the other six boxes while the other six boxes with abbreviations and different colors in the legend represent benchmarks.}
      \label{fig:Figure4}
\end{figure}

Finally, Figure \ref{fig:Figure4} presents the same simulation results as in Figure \ref{fig:Figure3} but assuming zero correlations between the factor series in each structure. Generally, the results seemingly exhibit a mixture effect of the two previous simulation results in Figures \ref{fig:Figure2} and \ref{fig:Figure3}. While $R^2$ statistics and RMSEs are increasing along the increase of SNR, the three congruence coefficients tend to remain the same. In particular, DSCA and DGICA have the same issue as seen in Figure \ref{fig:Figure2}, and the benefit of refitting can be observed again.

\subsection{Choice of initial rank selection}
\label{sse:result_bootstrap}

We consider several DGPs from Section \ref{sse:design} with various ranks and run the bootstrap procedure in Section \ref{ssse:initial_rank} to see if it can capture the correct initial ranks. We set the type as 1, and different sizes $(d,T)$ and SNR parameters $c$ are considered while $2K=100$ is fixed. We do not consider the case of growing $K$ since the rotational bootstrap is performed for each block, and gathering multiple data blocks does not help to estimate initial ranks. Then for all SNRs described above, we generate the DGPs along the combinations of ranks as follows:
\begin{itemize}
    \item[R.] $(r_J,r_G) = (1,1),(1,2),(1,3),(2,2),(1,4),(2,3),(2,4),(3,3),(3,4),(4,4)$.
\end{itemize}
Note that along the increase of $r_J$ and $r_G$, the corresponding transition matrices $\bar{\boldsymbol{\Phi}}$ and $\tilde{\boldsymbol{\Phi}}$ are adjusted by preserving \eqref{e:riccati_joint} and \eqref{e:riccati_group} with $\bar{\boldsymbol{\Psi}}$ and $\tilde{\boldsymbol{\Psi}}$ in \eqref{e:covariance_example} and fixed $\sigma_{\xi}$. Accordingly, we run the rotational bootstrap to evaluate the performance of the initial rank selection. For the combinations in R above, the initial ranks are $r=r_J+r_G=2,3,4,4,5,5,6,6,7,8$.

\begin{figure}[]
     \centering
     \includegraphics[width=1\textwidth,height=0.8\textheight]{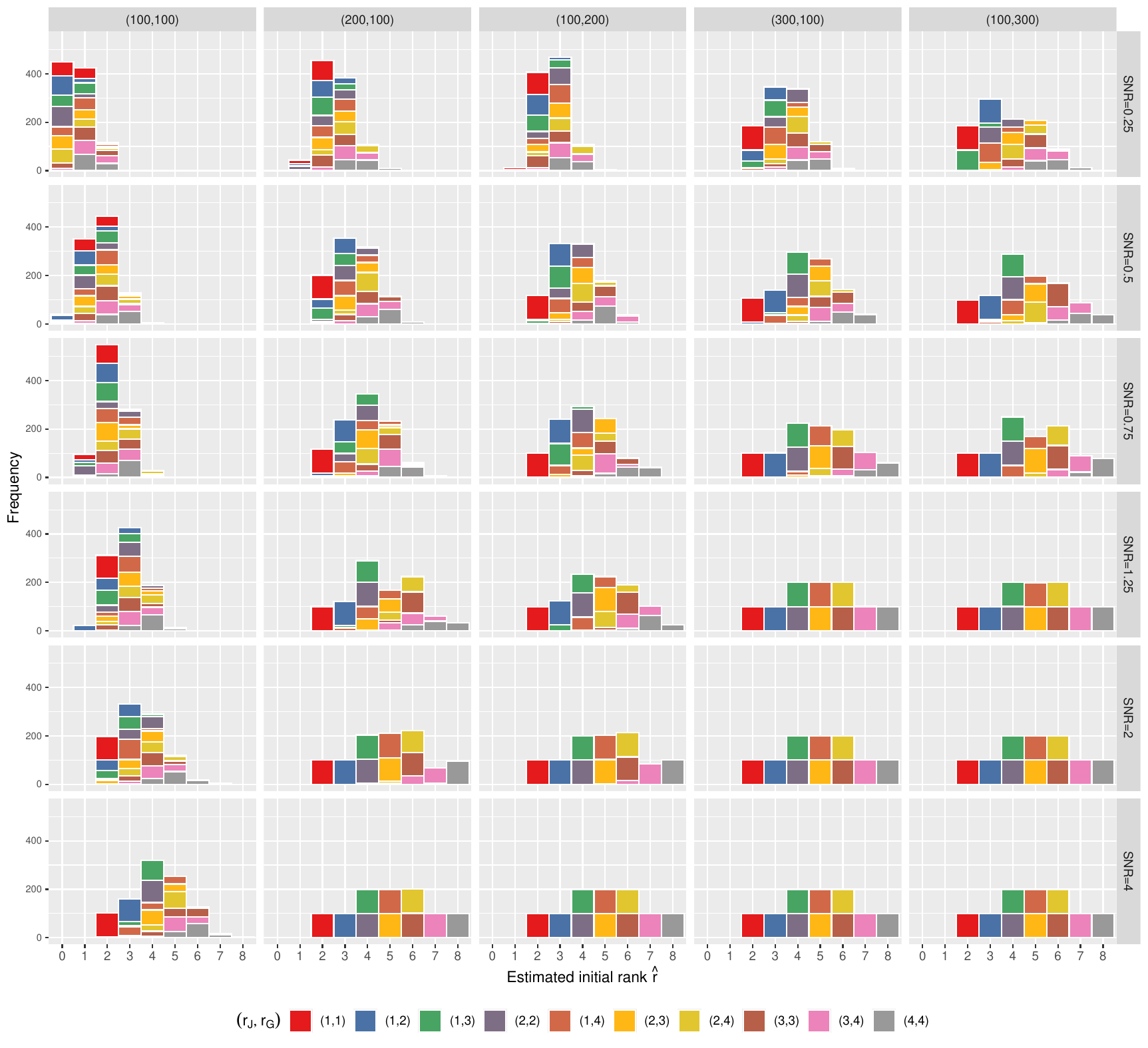}
     \caption{The frequencies of the estimated initial ranks are presented, with the expectation that as more ranks, denoted as $r$, are correctly estimated, the bars corresponding to those ranks should be positioned at $r = r_J + r_G$. Different colors and numbers in the parentheses in the legend represent different combinations of given $r_J$ and $r_G$. We employ SNR parameters ranging from $c=0.25$ to $c=4$, as described in Section \ref{sse:design}. Each column represents different problem sizes denoted as $(d,T)$. For all cases, we keep $2K$ fixed at 100, resulting in 100 replications for each setting.}
      \label{fig:Figure5}
\end{figure}

Figure \ref{fig:Figure5} shows several interesting results. First of all, the rank combinations used in Section \ref{sse:result_main} are purple, orange, and dark brown bars, respectively. These are usually correctly estimated for $(d,T)=(200,100)$ or $(d,T)=(100,200)$ when the SNR parameter is 1.25 or higher. This is also true for the rest of the higher initial ranks when $d$ or $T$ is increased to 300. This indicates that the large problem size seems advantageous in the same way the higher dimensions help in the usual factor models.

Furthermore, for the same initial total ranks, parameters and fixed problem sizes, the case where the number of joint factor series is equal to the number of group individual factors is more favorable than the case where the latter is larger. For example, the algorithm is more likely to choose the correct initial ranks in the case $(r_J,r_G)=(3,3)$ than the case $(r_J,r_G)=(2,4)$. Hence, larger ranks in the joint structure is favorable for selecting the correct ranks rather than larger ranks in the group individual structures.

\section{Data application}
\label{se:data_app}

\subsection{Autism Brain Imaging Data Exchange preprocessed data}
\label{sse:ABIDE}

In this section, we consider a large pre-processed dataset from the Autism Brain Imaging Data Exchange \citep[ABIDE preprocessed, e.g.,][]{craddock:2013}. ABIDE is an international collection of R-fMRI data from 20 site locations. The data consists of two different groups. Subjects with autism spectrum disorder (ASD) are marked as group 1. The rest of the subjects are from the control group, denoted as group 2. Group 1 consists of 474 males and 65 females and group 2 contains 474 males and 99 females. The average ages for the two groups are 17.01 and 17.08 years, respectively. The dataset from ABIDE is known to have been preprocessed by five teams with their preferred methods.

Among 1112 subjects, we work with 505 subjects from the ASD group and 530 subjects from the control group. We choose only the subjects whose profiles are identifiable from the separate phenotype file, and the missing observations of the subjects do not occur across all variables. However, some variables in some subjects still contain missing values. This dataset consists of an extensive array of BOLD signals. The brain atlas of \cite{dosenbach:2010} contains 160 ROIs. The ROIs belong to one of six brain regions, Default network, Frontoparietal network, Cingulo-Opercular network, Sensorimotor network, Cerebellum, and Occipital lobe. The numbers of ROIs in each network are 34, 21, 32, 33, 18, and 22, respectively. In our analysis, the average of the sample lengths over 1035 subjects is 193 time points. Each of 160 ROI labels consists of a combination of the network and the number. More specifically, the network means the brain region where the ROI is measured. In the illustration in Section \ref{sse:result_application}, different colors are used for representing the same unique regions.

\subsection{Connection to VAR-type model}
\label{sse:VARMA}

In this section, we briefly introduce the network construction by converting DFMs into VAR-type models. This will be useful in thinking about temporal and contemporaneous effects in the considered models. Consider the DFM in \eqref{e:augmented_factor_1}--\eqref{e:augmented_factor_2}. From \eqref{e:filter_equation}, we assume that the joint and group individual latent factor series follow VAR(1), $\bar{F}_{t}^{(k)}=\bar{\mathbf{\Psi}}_k\bar{F}_{t-1}^{(k)} + \bar{\eta}_{t}^{(k)}$ and $\tilde{F}_{t}^{(k)}=\tilde{\mathbf{\Psi}}_k\bar{F}_{t-1}^{(k)} + \tilde{\eta}_{t}^{(k)}$, respectively. Then, one can rewrite the DFM as 
\begin{eqnarray*}
    X_{t}^{(k)} 
    &=& \mathbf{B}_{g}F_{t}^{(k)} + E_{t}^{(k)} 
    = \begin{bmatrix} \bar{\mathbf{B}} & \tilde{\mathbf{B}}_{g} \end{bmatrix}
    \begin{bmatrix} \bar{F}_{t}^{(k)} \\ \tilde{F}_{t}^{(k)} \end{bmatrix} + E_{t}^{(k)} \\ 
    &=& \begin{bmatrix} \bar{\mathbf{B}} & \tilde{\mathbf{B}}_{g} \end{bmatrix}
    \left( \begin{pmatrix}
    \bar{\mathbf{\Psi}}_k & \boldsymbol{0}_{r_{J} \times r_{G}} \\
    \boldsymbol{0}_{r_{G} \times r_{J}} & \tilde{\mathbf{\Psi}}_k
    \end{pmatrix} \begin{bmatrix} \bar{F}_{t-1}^{(k)} \\ \tilde{F}_{t-1}^{(k)} \end{bmatrix} + \begin{bmatrix} \bar{\eta}_{t}^{(k)} \\ \tilde{\eta}_{t}^{(k)} \end{bmatrix} \right) + E_{t}^{(k)} \\
    &=:& \mathbf{B}_{g} \left( \mathbf{\Psi}_k F_{t-1}^{(k)} +\eta_{t}^{(k)} \right) + E_{t}^{(k)}.
\end{eqnarray*}
By using the fact that $F_{t-1}^{(k)} = (\mathbf{B}_{g}'\mathbf{B}_{g})^{-1}\mathbf{B}_{g}'(X_{t-1}^{(k)}-E_{t-1}^{(k)})$ where
\begin{equation*}
    (\mathbf{B}_{g}'\mathbf{B}_{g})^{-1} 
    = \begin{pmatrix}
    (\bar{\mathbf{B}}'\bar{\mathbf{B}})^{-1} & \boldsymbol{0}_{r_{J} \times r_{G}} \\
    \boldsymbol{0}_{r_{G} \times r_{J}} & (\tilde{\mathbf{B}}_g'\tilde{\mathbf{B}}_g)^{-1}
    \end{pmatrix},
\end{equation*}
$X_{t}^{(k)}$ can be expressed further as
\begin{eqnarray}
    X_{t}^{(k)} 
    &=& \left( \bar{\mathbf{B}}\bar{\mathbf{\Psi}}_k(\bar{\mathbf{B}}'\bar{\mathbf{B}})^{-1}\bar{\mathbf{B}}' + \tilde{\mathbf{B}}_g\tilde{\mathbf{\Psi}}_k(\tilde{\mathbf{B}}_g'\tilde{\mathbf{B}}_g)^{-1}\tilde{\mathbf{B}}_g' \right) X_{t-1}^{(k)} + \zeta_{t}^{(k)} \label{e:VARMA_trans1} \\
    &=:& \left( \bar{\mathbf{\Theta}}_{k} + \tilde{\mathbf{\Theta}}_{k} \right) X_{t-1}^{(k)} + \zeta_{t}^{(k)},\quad\ k=1,\ldots,2K, \label{e:VARMA_trans2}
\end{eqnarray}
where $\zeta_{t}^{(k)}$ is a centered noise with covariance matrix $\boldsymbol{\Sigma}_{\zeta,k}$, which is given by
\begin{equation}\label{e:VARMA_covariance}
    \boldsymbol{\Sigma}_{\zeta,k} = \left( \bar{\mathbf{\Theta}}_{k} + \tilde{\mathbf{\Theta}}_{k} \right)\left(\mathbf{I}_d + \boldsymbol{\Sigma}_{E,k}\right) + \bar{\mathbf{B}} \boldsymbol{\Sigma}_{\bar{\eta},k} \bar{\mathbf{B}}' + \tilde{\mathbf{B}}_g \boldsymbol{\Sigma}_{\tilde{\eta},k} \tilde{\mathbf{B}}_g',
\end{equation}
and $\bar{\mathbf{\Theta}}_{k} = \bar{\mathbf{B}}\bar{\mathbf{\Psi}}_k(\bar{\mathbf{B}}'\bar{\mathbf{B}})^{-1}\bar{\mathbf{B}}', \tilde{\mathbf{\Theta}}_{k}=\tilde{\mathbf{B}}_g\tilde{\mathbf{\Psi}}_k(\tilde{\mathbf{B}}_g'\tilde{\mathbf{B}}_g)^{-1}\tilde{\mathbf{B}}_g'$. Hence, for each subject $k$, the dynamics is defined by the two transition matrices $\bar{\mathbf{\Theta}}_{k},\tilde{\mathbf{\Theta}}_{k}$ in \eqref{e:VARMA_trans1}--\eqref{e:VARMA_trans2} and the error covariance matrix $\boldsymbol{\Sigma}_{\zeta,k}$ in \eqref{e:VARMA_covariance}. The equation \eqref{e:VARMA_trans2} is a high-dimensional VAR(1) model. Note that, unlike the high-dimensional VAR models where the sparsity on the transition matrices is imposed, the matrices $\bar{\mathbf{\Theta}}_{k},\tilde{\mathbf{\Theta}}_{k}$ are not sparse objects in general but have reduced rank. The model is akin to reduced-rank VAR since the rank of the transition matrix constructed by the sum of low-rank matrices cannot exceed the sum of two ranks. In the network context, the VAR transition matrix and covariance of the noise are called the directed network and contemporaneous network, respectively. From this perspective, we examine in the next section the differences between the two groups by comparing how their model network connectivities differ.

\subsection{Application results}
\label{sse:result_application}

In this section, we present estimation results. Some figures discussed below are placed in Section \ref{ap:fig_application} of the supplemental material. Figure \ref{fig:FigureS1} presents the results of the initial rank selection. Different colors, which are the same as used in quality assessment (QA) measures at the website \cite{abide:2016}, represent data collected from 20 different sites. The estimated initial ranks are divided into three regions: the left with lower ranks, the middle with intermediate ranks, and the right with higher ranks. The distribution of initially estimated ranks for group 1 closely resembles that of group 2. The data-collecting sites are more influential to the initial ranks, which suggests preprocessing differences, rather than the group differences. We expect that GRIDY framework can further facilitate the study of heterogeneity in R-fMRI across different multi-site differences \cite[e.g.,][]{abraham:2017,wang:2019} in future research.

We exclude data where the initially estimated ranks are zero or exceed 15. As it can be seen through the gray dashed line in Figure \ref{fig:FigureS1}, approximately 40\% of the subjects are retained for the remaining analysis. Consequently, the majority of the estimated ranks are 2. Interestingly, although the AJIVE result for the joint rank selection is 2, the ranks for the remaining individual structures within each group, as produced by AJIVE, are mostly 1. This indicates that AJIVE detects systematic variation within each subject even after factoring out the joint structures. Therefore, for the remaining analysis, we set the joint structure as rank 2 and model the group individual structures with a rank of 1 for each group. We also exclude subjects whose ranks for the group individual structures are zero for the rest of the procedure. Thus, 290 subjects are considered for the rest of the analysis, and each group has 157 and 137 subjects, respectively.

Next, we consider the estimates of the loadings matrices, covariance matrices, and the VAR transition matrices of the factor series separately. From the simulation in Section \ref{se:illustrative}, the proposed GRIDY and the two benchmarks SCA-P and GICA have shown better performance than the others, making it difficult to conclude which method is the best. Hence, we consider these three estimation methods. Figure \ref{fig:FigureS2} displays the estimated loadings matrices from the three methods, with each method presented in a separate column. All variables are reordered based on their networks, with each of the 6 networks represented by its own color. The first two rows and the second two rows represent the joint and group individual structures, respectively. Interestingly, while the landscape of the estimated loadings matrices in the group individual structures looks similar, one can notice both similarities and differences in the joint structures. For example, the values of the estimated loadings from each method in the group individual structures are very similar, except for their signs. However, while the estimates of the first and second loadings matrices from GRIDY resemble those of the second and first loadings matrices from GICA, the estimates of the loadings matrices from SCA-P do not exhibit similar patterns to those from the other methods.

In terms of the factor series, Figure \ref{fig:FigureS3} presents the sample covariances and VAR transition matrices of the factor series, estimated from the three methods. Interestingly, in the joint structure, the entries of the sample covariance matrices and the transition matrices exhibit considerable diversity across the methods. In particular, the off-diagonal entries of the sample covariance, indicating the covariance between the factors, differ noticeably among the methods. This discrepancy corresponds to the large range of the off-diagonal entries, which are often linked to the existence of Granger causality \cite[e.g., Chapter 2.3 in][]{lutkepohl2005new}. Notably, from the estimated loadings in Figure \ref{fig:FigureS2}, we observed that the two estimated loadings matrices from GRIDY look similar to those from GICA. The estimates for the factor series from the two methods are also similar to each other in their absolute scales, except that the orders are switched in the current figure.

\begin{figure}[]
     \centering
     \includegraphics[width=1\textwidth,height=0.6\textheight]{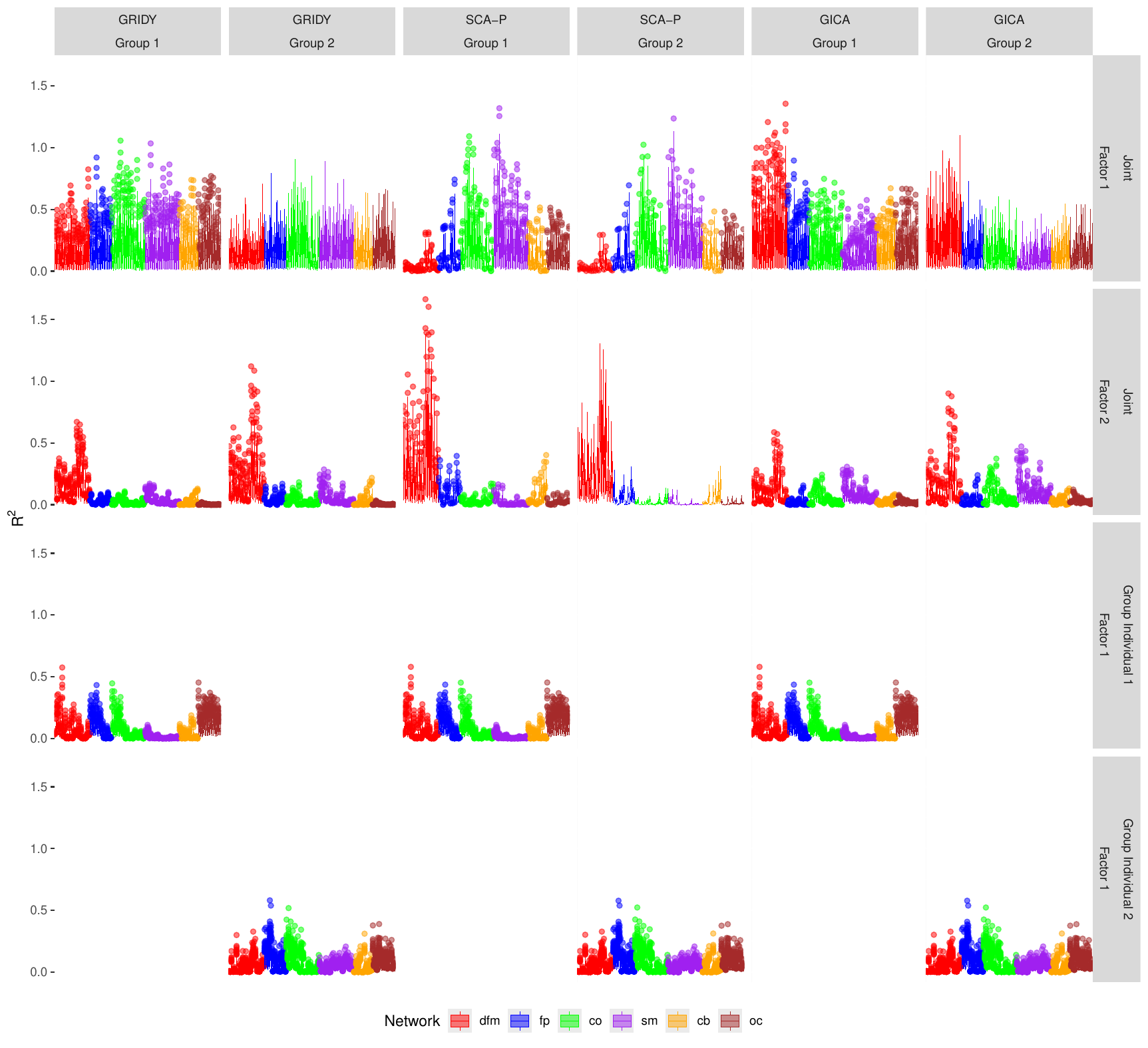}
     \caption{$R^2$ statistics for the estimated factors against each variable are presented across all subjects, computed from the estimates of each method (column-wise). The ROIs are reordered based on the network. The same colors of the labels indicate that the corresponding ROIs are known to belong to the same network, representing the Default network (red), Frontoparietal network (blue), Cingulo-Opercular network (green), Sensorimotor network (purple), Cerebellum (orange), and Occipital lobe (brown), respectively. The results from the ASD group and the control group are displayed in the odd and even columns, respectively. The results from the joint components are shown in the top two rows. The panels in the third row of the left column and the fourth row of the right column display the results from group individual components for each group.}
      \label{fig:Figure6}
\end{figure}

In Figure \ref{fig:Figure6}, we present box plots of the $R^2$ statistics across ROIs obtained by regressing the univariate observations $X_{i,t}^{(k)}$ on a particular factor estimate, as discussed in \cite{stock:2002} and \cite{jungbacker:2015}, from the three estimation methods. These statistics indicate how much variability for each variable can be explained by each factor. Interestingly, between the ASD group (on the left column of each panel) and the control group (on the right column of each panel), the distributions of $R^2$ statistics in each structure are quite similar. More intriguingly, while we confirmed that the estimated loadings matrices and the estimates for the factor series in the joint structures from GRIDY and GICA were similar, their $R^2$ statistics look quite different in terms of their vertical scales. In particular, values of $R^2$ statistics from SCA-P and GICA often exceed 1, while the values from GRIDY tend to remain below 1. By the definition of this statistic, values are allowed to exceed 1. However, if we consider the $R^2$ statistics structure-wise, it is difficult for the value to exceed 1 in general. This implies that the contribution to fitness from each factor series can be canceled out by the large estimated correlation between the factors.

\begin{figure}[]
     \centering
     \includegraphics[width=1\textwidth,height=0.7\textheight]{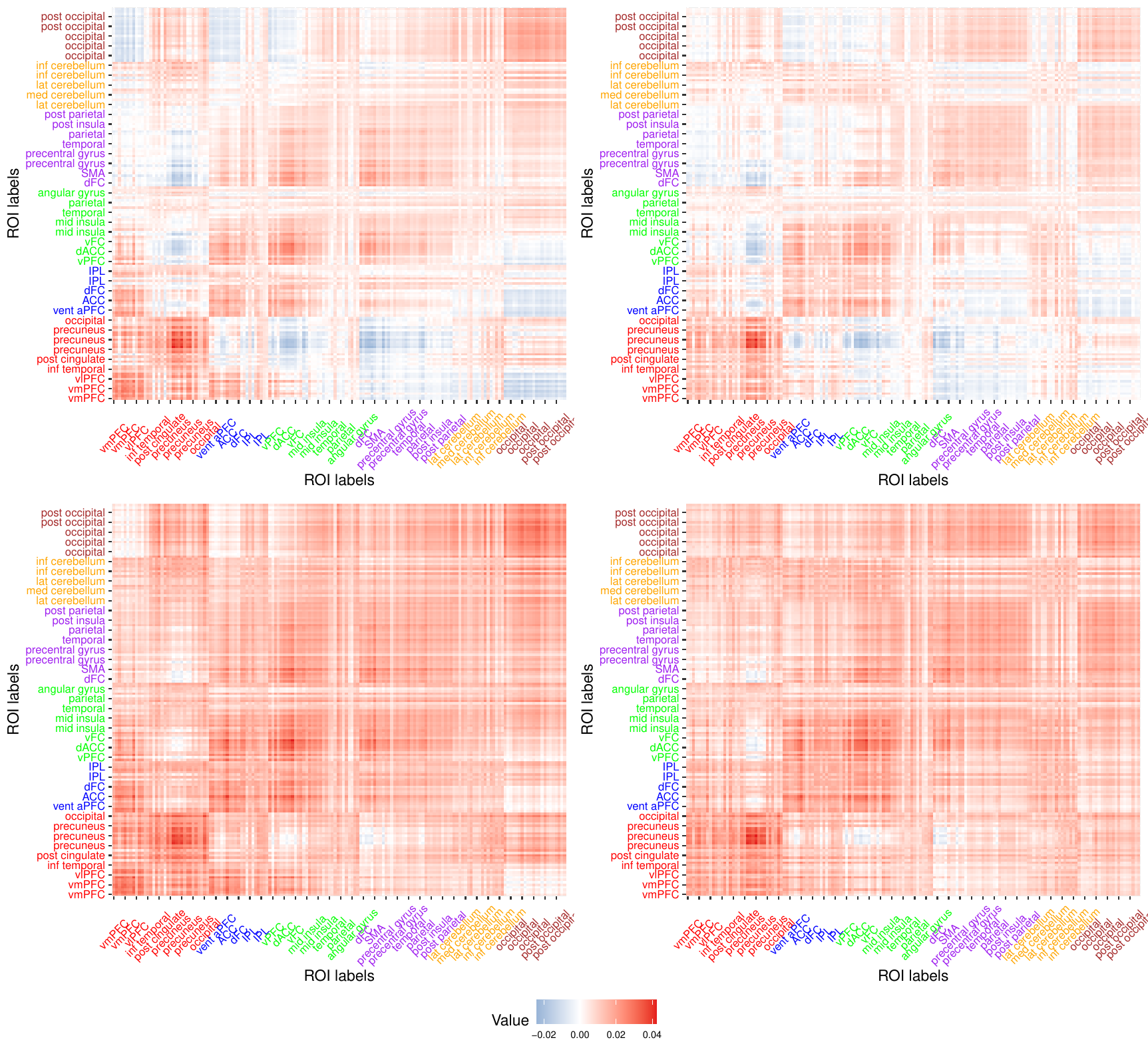}
     \caption{Heatmaps for the averages of estimated network matrices for the ASD group (left panel) and control group (right panel). Two plots in the top panel are directed networks and the two bottom plots are contemporaneous networks. The ROIs are reordered based on the network. Each color represents the Default network (red), Frontoparietal network (blue), Cingulo-Opercular network (green), Sensorimotor network (purple), Cerebellum (orange), and Occipital lobe (brown), respectively.}
      \label{fig:Figure7}
\end{figure}

Finally, the VAR connection explained in Section \ref{sse:VARMA} is considered to illustrate network structures. Figure \ref{fig:Figure7} summarizes the network structures across all subjects. The left panel presents the averages of the directed networks and the contemporaneous networks for group 1, while the results for group 2 are displayed in the right panel. The two heatmaps in the top panel represent the directed networks, and the two in the bottom panel represent contemporaneous networks. The ROIs in the networks are the same as in the previous figures. We can clearly observe distinct block structures in both the directed networks and contemporaneous networks. Interestingly, while the sizes and locations of the identified blocks in the directed networks are quite similar, the subpatterns within each block are quite different. Although the scales of the contemporaneous networks are relatively larger, they exhibit a similar tendency as the directed networks. On the contrary, although the contemporaneous network structures for both groups are also different, they exhibit relatively high similarity. Further study including confirmatory factor analysis is needed to link these observations with the clinical interpretations of the difference between the ASD group and the control group \cite[e.g.,][]{subbaraju:2017,easson:2019}.

\section{Discussion}
\label{se:discussion}

In this study, we propose a novel approach for group-level analysis of DFMs where the resulting integrated model is considered as the summation of joint components, group individual components, and the observation noise. Each signal structure is further decomposed into loadings matrices and latent factor series, with these series assumed to follow VAR models. Unlike the standard DFMs, our model allows different factor series within the same structure to be correlated. To handle such latent and correlated structures, we employ simultaneous component analysis. We conducted simulations to compare our modeling approach with several benchmark models suitable for the considered setting.  Using various estimation performance measures, the empirical evidence suggests that our approach outperforms the considered alternatives. To illustrate the practical application of our approach, ABIDE preprocessed dataset is used. This demonstrates how our approach can be employed to compare configurations and network structures between autism spectrum disorder and control groups.

There are several extensions and limitations of this work that could be addressed in the future. First, forecasting issues under the proposed model could be considered. In fact, we carried out simulations to assess forecasting performance, but among the methods considered, neither GRIDY nor other benchmarks were superior in terms of forecasting accuracy. Another practical issue to address is the case of missing observations. Indeed, missing values occurred even in our data application, and corresponding subjects had to be removed from the groups if the degree of missingness was severe. The two recent studies by \cite{bai:2021,xiong:2023} proposed ways to deal with missing data in large factor models that accommodate a range of missing patterns (beyond missingness at random) by relying on the factor structure. Whether and how those approaches could be extended to incorporate the dimension of subjects, in addition to the cross-sectional and temporal dimension, and to allow for both joint and individual factor structure is left for future work. Improvement of the proposed algorithm is also an interesting direction for future research. For example, one could consider adopting a more efficient algorithm for estimating the SCA-PF2 model, as developed by \cite{perros:2017}. This algorithm would prove indispensable in speeding up computation and improving memory efficiency, particularly as the number of subjects increases. Additionally, from the VAR representation discussed earlier, our model bears similarities to the multi-VAR model of \cite{fisher:2022} albeit with sparse VAR transition matrices. Lastly, a highly promising avenue for future research involves the consideration of multiple subgroups. In such cases, incorporating a partially-shared structure \cite[e.g.,][]{gaynanova:2019,prothero:2022} could allow for only some grouped subjects to possess shared structures when the number of groups exceeds two.

\section*{Acknowledgement}

Vladas Pipiras’s research was partially supported by the grants NSF DMS 1712966, DMS 2113662, and DMS 2134107.

\section*{Data availability statement}
The R code used in the simulations of Sections \ref{se:illustrative} and in the data analysis of Section \ref{se:data_app} is available on GitHub at \href{https://github.com/yk748/GRIDY}{https://github.com/yk748/GRIDY}. The ABIDE preprocessed data used in Section \ref{se:data_app} were derived from the following resource available in the public domain: \href{http://preprocessed-connectomes-project.org/abide/index.html}{http://preprocessed-connectomes-project.org/abide/index.html}. 

\section*{Conflict of interest}
The authors have declared no conflict of interest.

\clearpage

\clearpage
\small
\bibliography{multifac}

\appendix
\newpage
\setcounter{section}{0}
\setcounter{page}{1}
\setcounter{figure}{0}
\setcounter{equation}{0}
\renewcommand{\thesection}{S\arabic{section}}
\renewcommand{\thepage}{\arabic{page}}
\renewcommand{\thetable}{S\arabic{table}}
\renewcommand{\thefigure}{S\arabic{figure}}

\begin{center}
\textbf{\LARGE {Supplemental material to \\ ``Group integrative dynamic factor models with application to multiple subject brain connectivity}''}
\end{center}

\section{Summary of methods used in Section 4 with illustrative examples}
\label{ap:tab_summary}

\begin{table}[h]
\centering
\resizebox{\textwidth}{!}{%
\begin{tabular}{clll}
\cline{2-4}
                  & \multicolumn{3}{c}{Estimation procedure}                                                                                                                                                                   \\ \hline
Abbreviation      & \multicolumn{1}{c}{Segmentation of joint and group individual structures} & \multicolumn{1}{c}{Assumed factor models} & \multicolumn{1}{c}{Refitting}  \\ \hline
GRIDY             & Using AJIVE                                            & PARAFAC2                                  & Yes                                                 \\ \cline{1-1}
SCA-P             & Using AJIVE                                            & SCA-P                                     & Yes                                             \\
GICA              & Using AJIVE                                            & GICA                                      & Yes                                             \\ \cline{1-1}
DSCA              & Applying SCA-PF2 model twice                            & PARAFAC2                                  & Yes                                                  \\
DGICA             & Applying GICA model twice                               & GICA                                      & Yes                                             \\ \cline{1-1}
Unfitted\_SCA-PF2 & Using AJIVE                                            & PARAFAC2                                  & No                                                                            \\
Unfitted\_GICA    & Using AJIVE                                            & GICA                                      & No                                              \\ \hline
\end{tabular}%
}
\caption{Summary of similarities and differences of 7 models used in Section 4 with illustrative examples. The differences between estimation procedures lie in the choice of segmentation method, the assumed covariance structure of the factor models, and whether refitting is applied.}
\label{tab:summary}
\end{table}

\section{Additional figures for Section 5 with data application}
\label{ap:fig_application}

The three figures regarding the estimates of initial ranks, loadings matrices, and factor models for the Data application section (Section 5) are included in this section. The figures are discussed in Section \ref{sse:result_application}.

\begin{figure}[]
     \centering
     \includegraphics[width=1\textwidth,height=0.8\textheight]{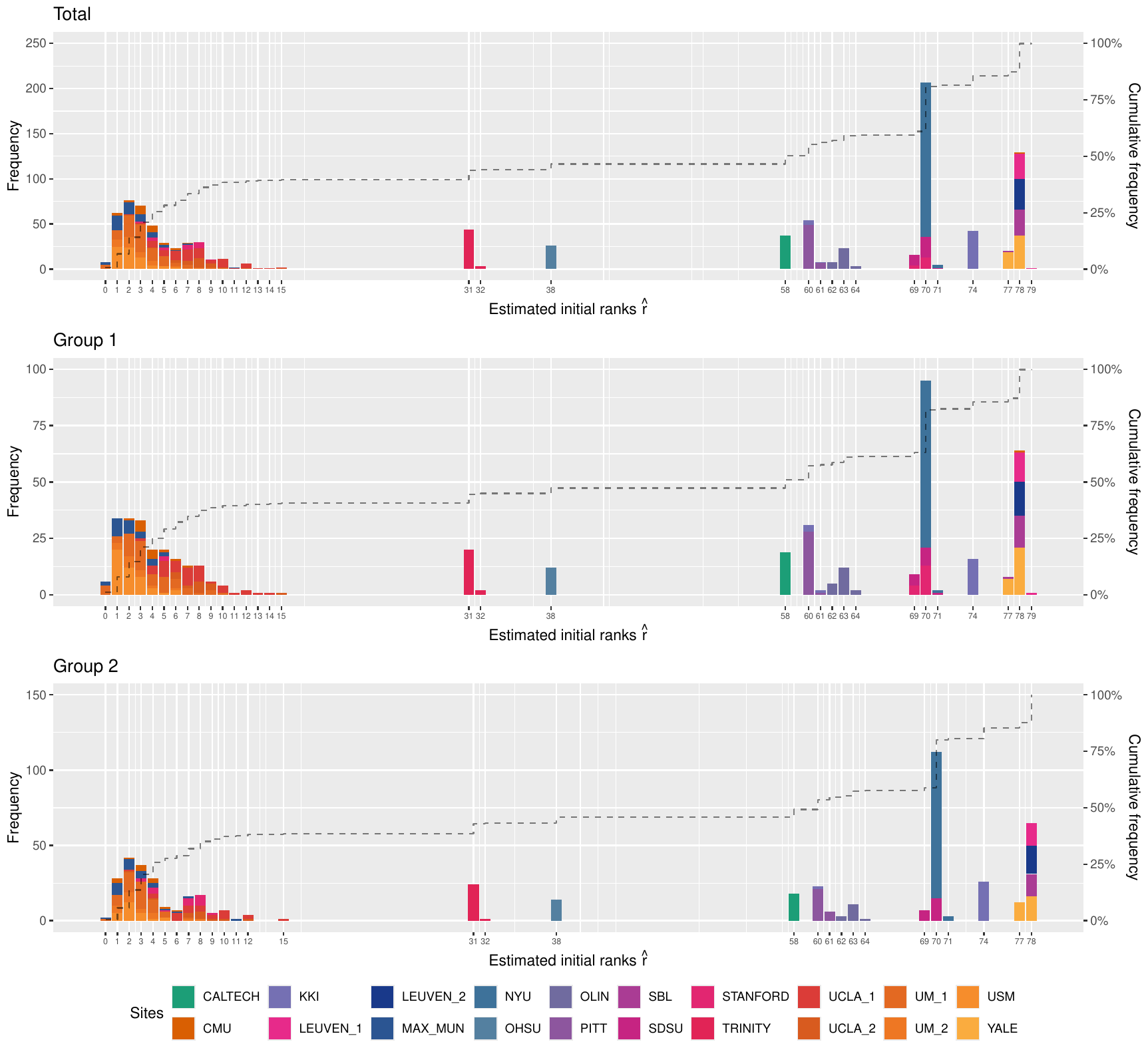}
     \caption{Initial rank estimation results. The abbreviations in the legend indicate the 20 different sites where the data were collected. The rank estimation result for the two groups is presented in the top panel. The estimation result for group 1 is in the middle panel and the result for group 2 is displayed in the bottom panel. The colored bars are estimated ranks for each site and the grey dashed line is for the cumulative sums of the colored bars for each estimated rank.}
      \label{fig:FigureS1}
\end{figure}

\begin{figure}[]
     \centering
     \includegraphics[width=1\textwidth,height=0.5\textheight]{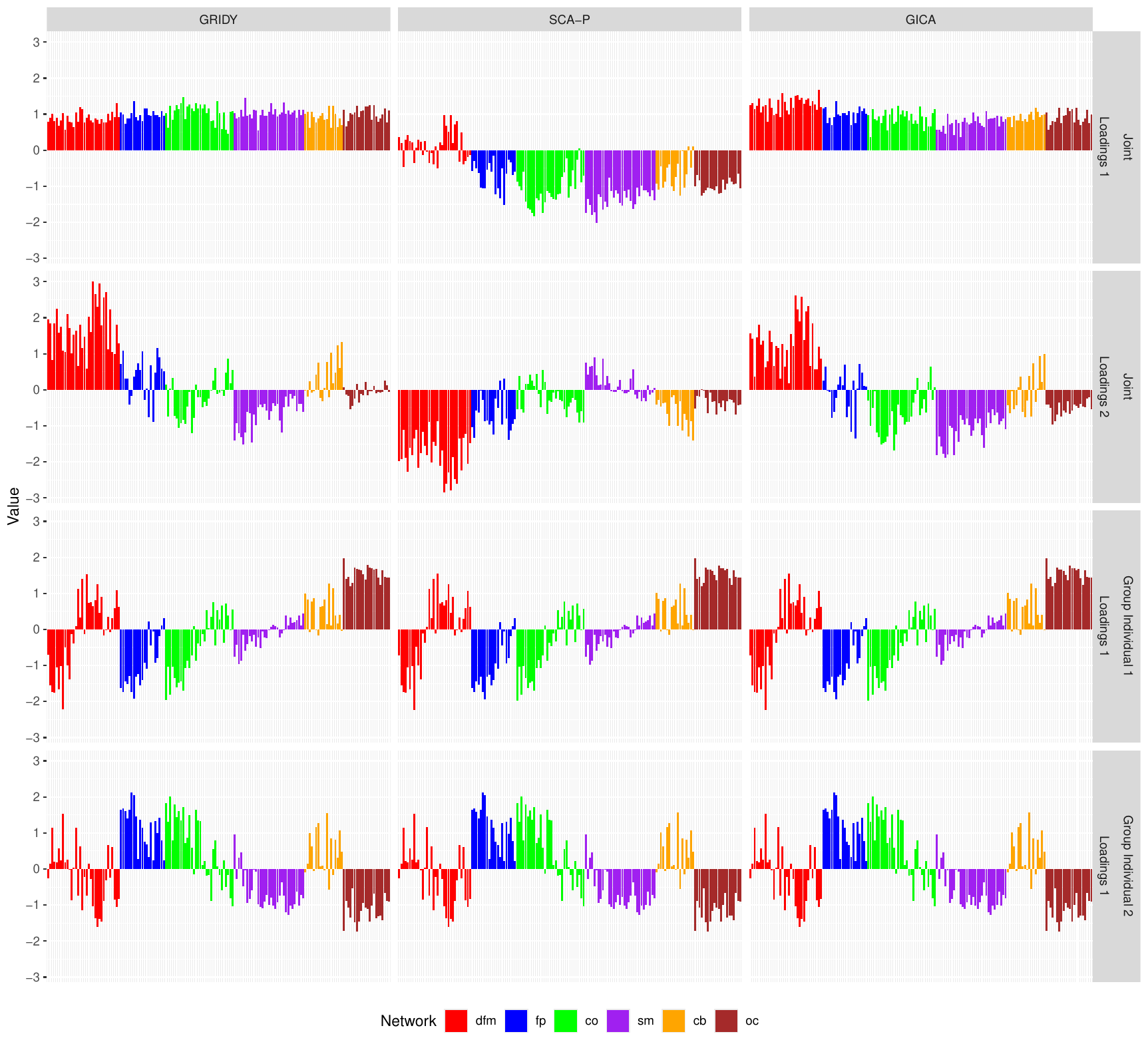}
     \caption{The estimated loadings matrices for the joint structures (top 2 rows), for the group individual structures in group 1 (third row), and for the group individual structures in group 2 (fourth row), estimated from three estimation methods (column-wise). The same colors of the labels indicate that the corresponding ROIs are known to belong to the same network. The colors used in the legend represent the Default network (red), Frontoparietal network (blue), Cingulo-Opercular network (green), Sensorimotor network (purple), Cerebellum (orange), and Occipital lobe (brown), respectively.}
      \label{fig:FigureS2}
\end{figure}

\begin{figure}[]
    \centering
    \includegraphics[width=1\textwidth,height=0.5\textheight]{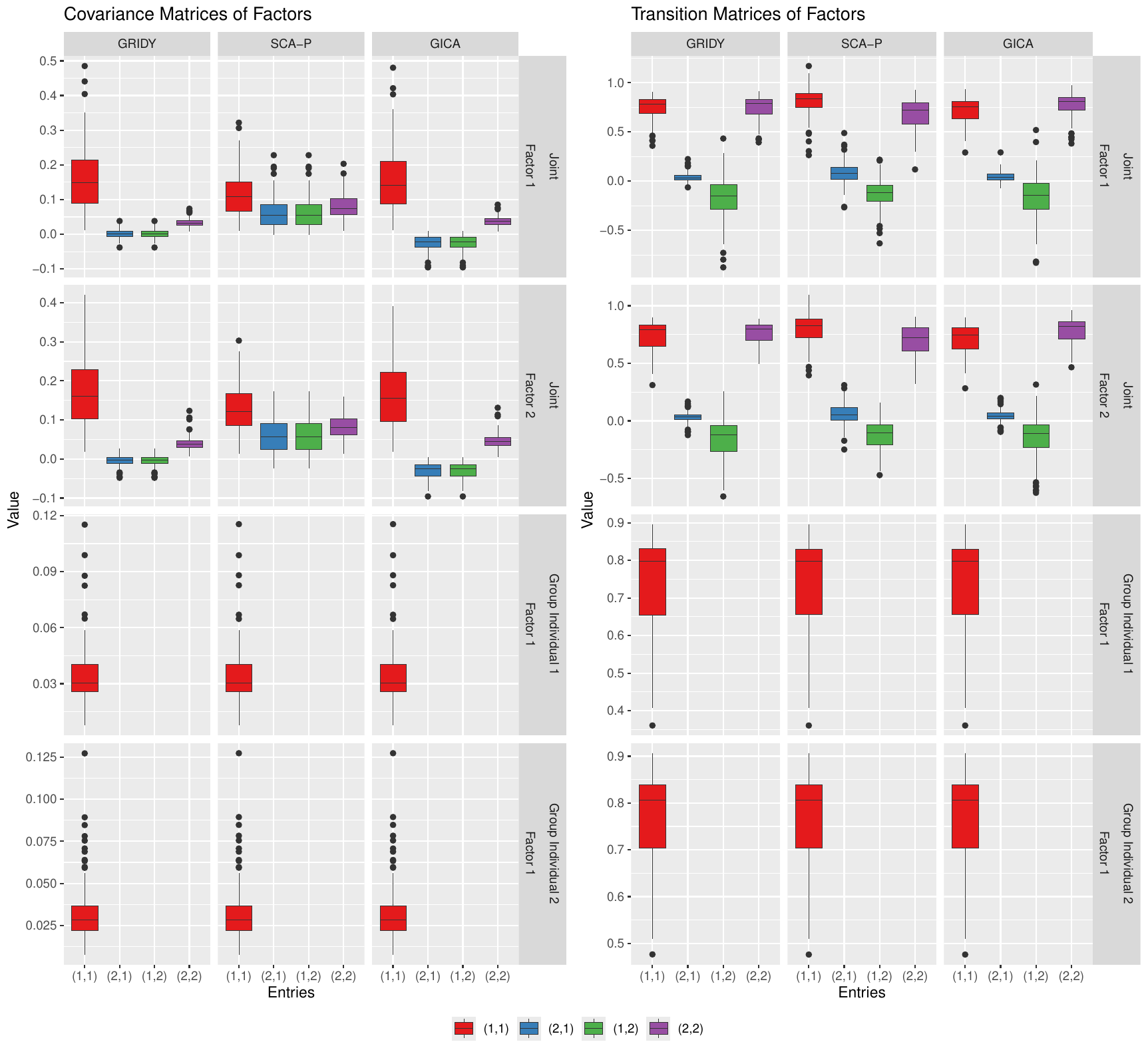}
    \caption{Estimated covariances of factor series (left) and VAR transition matrices (right) through the three fitting methods (columns in each panel). For each column, the first two rows concern the joint factor series and the last two columns concern the group individual factor series, each of the rows represents the estimates from the factor series belonging to each group. The same colors with coordinates in the legend are used for the estimates of the model parameters in the same coordinates.}
    \label{fig:FigureS3}
\end{figure}

\end{document}